\documentclass[amssymb,amsmath,prd,twocolumn,superscriptaddress,nofootinbib]{revtex4-1}

\usepackage{graphicx}
\usepackage{bm}
\usepackage{amssymb,amsmath}
\usepackage{mathrsfs}
\usepackage{latexsym}
\usepackage{color}
\usepackage[normalem]{ulem} 
\usepackage{dcolumn}
\usepackage[colorlinks=true,citecolor=blue,urlcolor=blue]{hyperref}
\usepackage[usenames,dvipsnames]{xcolor}

\allowdisplaybreaks

\newcommand{\CCA}{\affiliation{Center for Computational Astrophysics, Flatiron Institute, 162 5th Ave, New York, NY 10010}}
\newcommand{\RIT}{\affiliation{Center for Computational Relativity and Gravitation, Rochester Institute of Technology, Rochester, New York 14623, USA}}
\newcommand{\GA}{\affiliation{Center for Relativistic Astrophysics, Georgia Institute of Technology, Atlanta, Georgia 30332, USA}}
\newcommand{\AEI}{\affiliation{Max Planck Institute for Gravitational Physics (Albert Einstein Institute), Am M\"uhlenberg 1, Potsdam 14476, Germany}}
\newcommand{\MIT}{\affiliation{LIGO Laboratory and Kavli Institute for Astrophysics and Space Research, Massachusetts Institute of Technology, Cambridge, Massachusetts 02139, USA}}
\newcommand{\Australia}{\affiliation{Monash Centre for Astrophysics, School of Physics and Astronomy, Monash University, VIC 3800, Australia}}
\newcommand{\OzGrav}{\affiliation{OzGrav: The ARC Centre of Excellence for Gravitational-Wave Discovery, Clayton, VIC 3800, Australia }}
\newcommand{\Cardiff}{\affiliation{Gravity Exploration Institute, School of Physics and Astronomy, Cardiff University, The Parade, Cardiff CF24 3AA, UK}}
\newcommand{\AEIHannover}{\affiliation{Max Planck  Institute for Gravitational Physics (Albert Einstein Institute), Callinstr.~38, 30167 Hannover, Germany}}
\newcommand{\UniHannover}{\affiliation{Leibniz Universit\"at Hannover, D-30167 Hannover, Germany}}
\newcommand{\Manchester}{\affiliation{University of Manchester, Manchester, UK}}
\newcommand{\Caltech}{\affiliation{Theoretical Astrophysics 350-17, California Institute of Technology, Pasadena, CA 91125, USA}}
\newcommand{\Cornell}{\affiliation{Cornell Center for Astrophysics and Planetary Science, Cornell University, Ithaca, New York 14853, USA}}
\newcommand{\CITA}{\affiliation{Canadian Institute for Theoretical Astrophysics, 60 St.~George Street, University of Toronto, Toronto, ON M5S 3H8, Canada}}
\newcommand{\JPL}{\affiliation{Jet Propulsion Laboratory, California Institute of Technology, Pasadena, CA 91109, USA}}
\newcommand{\TorontoPhysics}{\affiliation{Department of Physics 60 St.~George Street, University of Toronto, Toronto, ON M5S 3H8, Canada}}
\newcommand{\GWPAC}{\affiliation{Gravitational Wave Physics and Astronomy Center, California State University Fullerton, Fullerton, California 92834, USA}}

\definecolor{kcmagenta}{rgb}{0.54, 0.17, 0.88}
\definecolor{darkturquoise}{rgb}{0.0, 0.81, 0.82}
\definecolor{chgreen}{rgb}{0.03,0.47,0.19}

\newcommand{\IMRP}{{\tt IMRPhenomPv2}}
\newcommand{\IMRPD}{{\tt IMRPhenomD}}
\newcommand{\IMRPHM}{{\tt IMRPhenomHM}}
\newcommand{\SEOBP}{{\tt SEOBNRv3}}
\newcommand{\SEOBA}{{\tt SEOBNRv4}}
\newcommand{\SEOBHM}{{\tt SEOBNRv4HM}}
\newcommand{\NRSur}{{\tt NRSur7dq2}}
\newcommand{\lm}{{\ell,m}}

\begin{document}

\title{On the properties of the massive binary black hole merger GW170729}

\author{Katerina Chatziioannou}
\CCA
\author{Roberto Cotesta}
\AEI
\author{Sudarshan Ghonge}
\GA
\author{Jacob Lange}
\RIT
\author{Ken K.-Y. Ng}
\MIT
\author{Juan Calder\'{o}n Bustillo}
\Australia
\OzGrav
\author{James Clark}
\GA
\author{Carl-Johan Haster}
\MIT
\author{Sebastian Khan}
\AEIHannover
\UniHannover
\author{Michael P\"{u}rrer}
\AEI
\author{Vivien Raymond}
\Cardiff
\author{Salvatore Vitale}
\MIT

\author{Nousha Afshari}
\GWPAC
\author{Stanislav Babak}
\affiliation{APC, Univ Paris Diderot, CNRS/IN2P3, CEA/Irfu, Obs de Paris, Sorbonne Paris Cit\'e, France}
\affiliation{Moscow Institute of Physics and Technology, Dolgoprudny, Moscow region, Russia}
\author{Kevin Barkett}
\Caltech
\author{Jonathan Blackman}
\Caltech
\author{Alejandro Boh\'e}
\affiliation{Centre National d'Etudes Spatiales, 18 av Edouard Belin, 31400 Toulouse, France}
\author{Michael Boyle}
\Cornell
\author{Alessandra Buonanno}
\AEI
\affiliation{Department of Physics, University of Maryland, College Park, Maryland 20742, USA}
\author{Manuela Campanelli}
\RIT
\author{Gregorio Carullo}
\affiliation{Dipartimento di Fisica ``Enrico Fermi'', Universit\`a di Pisa, Pisa I-56127, Italy}
\affiliation{INFN sezione di Pisa, Pisa I-56127, Italy}
\author{Tony Chu}
\CITA
\author{Eric Flynn}
\affiliation{Gravitational Wave Physics and Astronomy Center, California State University Fullerton, Fullerton, California 92834, USA}
\author{Heather Fong}
\TorontoPhysics
\CITA
\author{Alyssa Garcia}
\GWPAC
\author{Matthew Giesler}
\Caltech
\author{Maria Haney}
\affiliation{Physik-Institut, Universit\"at Z\"urich, Winterthurerstrasse 190, 8057 Z\"urich, Switzerland}
\author{Mark Hannam}
\Cardiff
\author{Ian Harry}
\affiliation{Institute of Cosmology and Gravitation, University of Portsmouth, Dennis Sciama Building, Burnaby Road, Portsmouth, PO1 3FX, United Kingdom}
\AEI
\author{James Healy}
\RIT
\author{Daniel Hemberger}
\Caltech
\author{Ian Hinder}
\AEI
\Manchester{}
\author{Karan Jani}
\GA
\author{Bhavesh Khamersa}
\GA
\author{Lawrence E. Kidder}
\Cornell
\author{Prayush Kumar}
\Cornell
\author{Pablo Laguna}
\GA
\author{Carlos O. Lousto}
\RIT
\author{Geoffrey Lovelace}
\GWPAC
\author{Tyson B. Littenberg}
\affiliation{NASA Marshall Space Flight Center, Huntsville, Alabama 35812, USA}
\author{Lionel London}
\MIT
\author{Margaret Millhouse}
\affiliation{OzGrav, School of Physics, University of Melbourne, Parkville, Victoria 3010, Australia}
\author{Laura K. Nuttall}
\affiliation{Institute of Cosmology and Gravitation, University of Portsmouth, Dennis Sciama Building, Burnaby Road, Portsmouth, PO1 3FX, United Kingdom}
\author{Frank Ohme}
\AEIHannover
\UniHannover
\author{Richard O'Shaughnessy}
\RIT
\author{Serguei Ossokine}
\AEI
\author{Francesco Pannarale}
\affiliation{Dipartimento di Fisica, ``Sapienza'' Universit\`a di Roma and Sezione INFN Roma1, Piazzale Aldo Moro 5, 00185 Roma, Italy}
\author{Patricia Schmidt}
\affiliation{School of Physics and Astronomy and Institute for Gravitational Wave Astronomy, University of Birmingham, Edgbaston, Birmingham, B15 9TT, United Kingdom}
\author{Harald P. Pfeiffer}
\AEI
\author{Mark A. Scheel}
\Caltech
\author{Lijing Shao}
\affiliation{Kavli Institute for Astronomy and Astrophysics, Peking University, Beijing 100871, China}
\author{Deirdre Shoemaker}
\GA
\author{Bela Szilagyi}
\Caltech
\JPL
\author{Andrea Taracchini}
\AEI
\author{Saul A. Teukolsky}
\Cornell
\Caltech
\author{Yosef Zlochower}
\RIT

\date{\today}

\begin{abstract}

We present a detailed investigation into the properties of GW170729, the gravitational wave with the most massive and distant source confirmed to date. 
We employ an extensive set of waveform models, including new improved models that incorporate the effect of higher-order waveform modes which are particularly important for massive systems.
We find no indication of spin-precession, but the inclusion of higher-order modes in the models results in an improved estimate for the mass ratio of $(0.3-0.8)$ at the 90\% credible level. Our updated measurement excludes equal masses at that level.
We also find that models with higher-order modes lead to the data being more consistent with a smaller effective spin, with the probability that the effective spin is greater than zero being reduced from $99\%$ to $94\%$.
The 90\% credible interval for the effective spin parameter is now $(-0.01-0.50)$.
Additionally, the recovered signal-to-noise ratio increases by $\sim0.3$ units compared to analyses without higher-order modes;
the overall Bayes Factor in favor of the presence of higher-order modes in the data is 5.1:1.
We study the effect of common spin priors on the derived spin and mass measurements, and observe small shifts in the spins, while the masses remain unaffected.  
We argue that our conclusions are robust against systematic errors in the waveform models. 
We also compare the above waveform-based analysis which employs compact-binary waveform models to a more flexible wavelet- and chirplet-based analysis. 
We find consistency between the two, with overlaps of $\sim 0.9$, typical of what is expected from simulations of signals similar to GW170729, confirming that the data are well-described by the existing waveform models. 
Finally, we study the possibility that the primary component of GW170729 was the remnant of a past merger of two black holes and find this scenario
to be indistinguishable from the standard formation scenario.
\end{abstract}

\maketitle

\section{Introduction}

GW170729 was observed on July 29, 2017 by the Advanced LIGO~\cite{TheLIGOScientific:2014jea} twin detectors. 
Its detection was announced in~\cite{LIGOScientific:2018mvr} as part of GWTC-1, the gravitational-wave (GW) transient catalog of compact binary coalescences (CBCs)~\cite{GWOSC:GWTC}. 
As reported in~\cite{LIGOScientific:2018mvr}, GW170729 was emitted during the coalescence of two stellar-mass black holes (BH). 
It was observed during the offline analysis of the detection pipelines GstLAL~\cite{Messick:2016aqy} and PyCBC~\cite{Usman:2015kfa,Nitz_2017}, that search for signals from CBC events, as well as cWB~\cite{Klimenko:2015ypf}, a pipeline tuned to search for transient signals whose frequency increases with time. 

A number of reasons make GW170729 unique among the binary BHs (BBHs) presented in GWTC-1.
At a measured source-frame total mass of $\sim 85 M_{\odot}$ and a distance of $\sim 3$Gpc (median values), it is likely the most massive and distant BBH. 
Additionally, it is one of only two GW events that show evidence for nonzero spins with an effective spin of $(0.11-0.58)$ at the 90\% credible level~\cite{LIGOScientific:2018mvr}. 
Finally, it is the only event for which the more flexible, non-CBC-specific cWB search returns a lower false alarm rate than the CBC-specific GstLAL and PyCBC template-based searches. 
In spite of these, Ref.~\cite{LIGOScientific:2018jsj} concludes that GW170729 is consistent with the population of the other BBH detections
\footnote{An additional candidate claimed in~\cite{Zackay:2019tzo}, if confirmed, would also correspond to a binary with a non zero effective spin.}.

The fact that GW170729 is the most massive BBH found so far makes it a good candidate to observe the effects of higher-order waveform modes. 
The GW emission from BBHs can be described as a superposition of GW modes $h_\lm$ as $h\equiv h_+ - i h_{\times}=\sum_\lm    Y_\lm^{-2}(\iota,\varphi)h_\lm(t,\vec{\theta})$ \cite{Thorne:1980ru}. 
Here  $h_+$ and $h_\times$ are the two GW polarizations, $Y_\lm^{-2}$ denote spin-2 weighted spherical harmonics \cite{sharm}, which depend on the location $(\iota,\varphi)$ of the observer around the binary, while the modes $h_\lm$ depend on the masses and spins of the binary, denoted by $\vec{\theta}$. 
During most of the inspiral stage, $h$ is dominated by the quadrupole modes, $(\lm)=(2,\pm 2)$. 
The rest, known as higher-order modes, grow in strength during the merger and ringdown stages, their impact being larger for highly asymmetric and nearly edge-on binaries \cite{Pan:2011gk,Varma:2014jxa,Bustillo:2016gid,Pekowsky:2012sr,Varma:2016dnf,Bustillo:2015qty,Graff:2015bba}. 
Finally, for more massive BBH systems the inspiral emission moves out of the sensitive band of advanced detectors, while sensitivity to the merger-ringdown increases and so does the impact of higher-order modes \cite{VanDenBroeck:2006ar,Bustillo:2015qty,Graff:2015bba}.

Standard detection and parameter estimation of BBH events is usually performed using GW templates without the higher-order mode content of the signals~\cite{2018arXiv181205121M,DalCanton:2017ala,TheLIGOScientific:2016pea,LIGOScientific:2018mvr}. 
Reference~\cite{LIGOScientific:2018mvr} studied the fact that the event is recovered with higher significance by the flexible unmodeled, but less sensitive, search than the template-based searches~\cite{CalderonBustillo:2017skv}. 
By performing injections of signals without higher-order modes or spin-precession, it was argued that the difference in the measured significances is in fact not unlikely. 
It was shown that $\sim 4\%$ of the injected signals were recovered with a higher significance from cWB than PyCBC. 

At the same time, the presence of strong higher-order modes in the GW signal can potentially lead to biased parameter estimation if they are omitted in the waveform templates \cite{Littenberg:2012uj,Varma:2016dnf,Varma:2016dnf,Bustillo:2015qty,Graff:2015bba}. 
So far all reported events are consistent with nearly equal-mass, face-on BBHs, a fact that has prevented such biases, as shown in \cite{Abbott:2016wiq,Abbott:2016apu,Kumar:2018hml} for the case of events observed during the first observation run of Advanced LIGO. 
Even in this case, usage of models with higher-order modes can improve the accuracy of parameter estimation \cite{Littenberg:2012uj,Graff:2015bba,London:2017bcn,CalderonBustillo:2018zuq}. 
Consequently, a reanalysis of GW150914 and GW170104 events using models with higher-order modes obtains modestly tighter parameter constraints with respect to previous analyses \cite{Kumar:2018hml}.

In this paper we present a detailed investigation into the properties of GW170729. 
We carry out a parameter estimation analysis similar to the one in~\cite{LIGOScientific:2018mvr} in order to study the effect of spin-precession, higher-order modes, and spin priors on inferences drawn about GW170729. 
We make use of a more extended set of CBC waveform models belonging to three distinct waveform model families: phenomenological~\cite{Ajith:2007qp,Ajith:2009bn,Santamaria:2010yb,Husa:2015iqa,Khan:2015jqa,London:2017bcn,Hannam:2013oca}, effective-one-body~\cite{Taracchini:2013rva,Bohe:2016gbl,Pan:2013rra,Cotesta:2018fcv,Pan:2011gk,Babak:2016tgq}, and numerical relativity~\cite{2013PhRvL.111x1104M,2016CQGra..33t4001J,2017CQGra..34v4001H,2015PhRvL.115l1102B,2017PhRvD..95j4023B,2017PhRvD..96b4058B}. 
This set includes two new improved spin-aligned waveform models that include the effect of higher-order waveform modes.
We gauge the importance of a physical phenomenon, namely spin-precession and higher-order modes, by comparing posterior densities for various source parameters obtained through analyses using waveform models with and without that physical phenomenon included.

While we find no indication of spin-precession, higher-order modes have a distinct impact on the posterior density for various source parameters. 
We find that CBC waveform models that include higher-order modes result in posterior distributions for the mass ratio of the binary that are shifted away from unity, resulting in more support for unequal masses than originally concluded in ~\cite{LIGOScientific:2018mvr}. 
In particular, we find a highest probability density (HPD) interval of the mass ratio of $(0.3-0.8)$ at the 90\% credible level, while the corresponding upper limit of the HPD interval on the mass ratio without higher-order modes is $\sim 0.96$.
This improved measurement, obtained using waveform models that include more physical effects, shows that GW170729 is not consistent with the merger of two equal-mass BHs at the $90\%$ level.
At the same time, models with higher-order modes lead to marginally less support for positive effective spin $\chi_{\rm{eff}}$~\footnote{The effective spin parameter is defined as the sum of the mass-weighted projections of the component spins along the orbital angular momentum and it is conserved to at least the second post-Newtonian order~\cite{Racine:2008qv}.} and binary orientations where the orbital angular momentum points along or away from the line of sight. In particular, we find that the probability that the effective spin is positive is reduced from $99\%$ when higher-order modes are omitted to $94\%$ when they are included in the waveform models.
We obtain consistent results when we use various CBC waveform models from different waveform families to describe the data. 
We thus argue that our conclusions are robust against systematic errors in the waveform models.

As the source-frame mass of the more massive BH is close to the proposed mass upper limit due to pair instability\footnote{It has been suggested that pulsational pair instability supernovae will result in no BH remnants with masses above $\sim 50 M_{\odot}$ as the remnant is disrupted during the explosion, e.g.~\cite{Woosley_2017,Marchant:2018kun}.}~\cite{2002ApJ...567..532H} and the posterior for the binary mass ratio favors unequal masses, we further investigate the possibility of second generation (2g) merger~\cite{davide2g,Fishbach:2017dwv}.
In a 2g merger scenario, the primary BH is the remnant of an earlier BBH merger. 
As such, it is expected to be more massive than its companion in GW170729 resulting in unequal binary masses, and to have a relatively large spin magnitude.
We contrast this scenario to a first generation (1g) merger scenario which favors comparable component masses.
We reanalyze the data using two different priors tailored to 1g and 2g mergers and calculate the Bayes Factor (BF) of the 2g versus the 1g hypotheses.
We find a BF of 4.7:1(1.4:1) in favor of the 2g scenario when using waveforms with (without) higher-order modes. This value favors the 2g model, but not decisively so, in agreement with the results of~\cite{LIGOScientific:2018jsj}. This question has been earlier and independently addressed by Kimball et al.~\cite{Kimball:2019mfs} and our results are in agreement with their results.

Finally, we compare the signal reconstruction obtained with CBC waveform models to a morphology-independent reconstruction~\cite{Cornish:2014kda}.
We quantify the consistency between the CBC and the generic reconstruction by computing the noise-weighted overlap between the two. 
We find broad consistency between the two with overlap values typical of what is expected for this mass range and signal strength~\cite{TheLIGOScientific:2016uux}. 
This result is only minimally affected by the inclusion of higher-order modes. 

Posterior samples from all our analyses are available in~\cite{170729SamplesRelease}.
The rest of the paper presents our analysis and conclusions in detail. 
Section~\ref{analysis} describes the analysis we carry out including the CBC waveform models, the priors, and the generic analysis. 
Section~\ref{CBCresults} presents results derived under the CBC waveform models and posterior densities for the source parameters. 
Section~\ref{BWresults} gives the results of the generic analysis and how they compare to the CBC-specific analysis. 
Finally, Sec.~\ref{conclusions} presents our main conclusions.

\section{Analysis}
\label{analysis}

In this section we describe the details of the analysis we perform including the data, waveform models, and inference approaches we use. 
Our analysis follows closely and builds off of the work originally presented in~\cite{LIGOScientific:2018mvr}. 
We employ two complementary approaches: one is based on waveform models constructed specifically to describe compact binary coalescences, while the other uses a more flexible waveform model that can capture unexpected signal morphology. 
We describe both analyses in the following.

\subsection{Data and setup} 

We use the publicly available LIGO and VIRGO strain data for GW170729 from the Gravitational Wave Open Science Center~\cite{GWOSC,Vallisneri:2014vxa}. 
The LIGO strain data have been post-processed to subtract several sources of instrumental noise~\cite{Driggers:2018gii,Davis:2018yrz} and calibrated as described in~\cite{LIGOScientific:2018mvr}. 
In particular we analyze $4$s of strain data centered at the GW170729 trigger time.
The analysis covers a bandwidth from $f_{\rm low}=20$~Hz with the upper frequency cutoff set to $f_{\rm high}=1024$~Hz for waveform models without higher-order modes and $f_{\rm high}=2048$~Hz when higher-order modes are included. 
For masses typical of GW170729 (a detector-frame total mass of 120$M_{\odot}$ and mass ratio of 0.5) this upper frequency cutoff ensures that the analysis includes up to at least the $\ell=5$ ringdown harmonic of a compact binary merger, the highest frequency mode available in the waveform models we use. 

We assume that the noise in the three detectors is Gaussian and stationary. 
The power spectral density (PSD) of the noise is obtained from the same $4$s of on-source data with the technique described in~\cite{Littenberg:2014oda}. 
Specifically, a model consisting of a cubic spline and a number of Lorenzians is used to obtain posterior samples for the PSD from which a median PSD value is computed separately for each frequency bin. 
This median PSD is used in the estimation of the likelihood function in {\tt LALInference}, {\tt RIFT}, and {\tt BayesWave} and we use the same PSD as~\cite{LIGOScientific:2018mvr} that is publicly available in~\cite{GW170729PSD}.

\subsection{ CBC waveform models} 

\begin{table*}[]
\centering
  \begin{tabular}{c|ccccccccc}
    \hline \hline
Waveform Model         &&& Spin Dynamics &&& Modes ($\ell,|m|$) &&&  Algorithm   \\ \hline
{\tt IMRPhenomPv2}~\cite{Hannam:2013oca}   &&& Precessing &&& (2,2) &&& {\tt LALInference}\\
{\tt IMRPhenomD}~\cite{Husa:2015iqa,Khan:2015jqa}       &&& Aligned &&& (2,2) &&&  {\tt LALInference}\\
{\tt IMRPhenomHM}~\cite{London:2017bcn}    &&& Aligned &&& (2,2),(2,1),(3,3),(3,2),(4,4),(4,3) &&& {\tt LALInference}\\
{\tt SEOBNRv3}~\cite{2014PhRvD..89f1502T,2014PhRvD..89h4006P}           &&& Precessing &&& (2,2) (2,1)  &&&   {\tt LALInference}\\
{\tt SEOBNRv4}~\cite{Bohe:2016gbl}           &&& Aligned &&&(2,2) &&& {\tt LALInference}\\
{\tt SEOBNRv4HM}~\cite{Cotesta:2018fcv}     &&& Aligned &&& (2,2), (2,1), (3,3), (4,4), (5,5)  &&& {\tt LALInference}\\
NR    HM ~\cite{2013PhRvL.111x1104M,2016CQGra..33t4001J,2017CQGra..34v4001H}            &&& Aligned &&&   ($\ell \leq 4$,$|m|\leq\ell$) &&& {\tt RIFT}\\
NR/\NRSur{} HM ~\cite{2015PhRvL.115l1102B,2017PhRvD..95j4023B,2017PhRvD..96b4058B} &&& Aligned &&&  ($\ell \leq 4$,$|m|\leq\ell$) &&& {\tt RIFT}\\
NR  ~\cite{2013PhRvL.111x1104M,2016CQGra..33t4001J,2017CQGra..34v4001H}           &&& Aligned &&&  ($\ell=2$,$|m|\leq\ell$) &&&{\tt RIFT}\\
NR/\NRSur{} ~\cite{2015PhRvL.115l1102B,2017PhRvD..95j4023B,2017PhRvD..96b4058B} &&& Aligned &&&  ($\ell=2$,$|m|\leq\ell$)  &&& {\tt RIFT}\\
\hline
Wavelets~\cite{Cornish:2014kda}               &&& Flexible &&&  Flexible &&&{\tt BayesWave}\\
Chirplets~\cite{Millhouse:2018dgi}           &&& Flexible &&&  Flexible  &&& {\tt BayesWave}\\
\end{tabular}
\caption{List of waveform models we use to model the GW signal. 
The second and third columns indicate the spin dynamics and higher-order content $(\ell, |m|)$  of each model in the coprecessing frame respectively. 
The fourth column gives the algorithm we use with each model. 
The bottom horizontal line separates the CBC-specific models and the morphology-independent models.
}
\label{table:pe-table}
\end{table*}

The top half of Table~\ref{table:pe-table} lists the CBC waveform models we use; these models describe the inspiral, merger, and ringdown signal from the coalescence of two BHs as predicted by General Relativity (GR). 
All CBC waveforms we use can be divided into three main families: (i) phenomenological models ({\tt IMRPhenom}), effective-one-body models ({\tt SEOBNR}), and numerical relativity (NR). 
The first family is based on results of post-Newtonian theory~\cite{Blanchet:2014zz} to compute the inspiral phase and a phenomenological approach to describe the merger, aided by calibration to EOB-NR hybrid waveforms~\cite{Hannam:2013oca,Husa:2015iqa,Khan:2015jqa}.
The second family uses the effective-one-body approach~\cite{Buonanno:1998gg,Buonanno:2000ef}, which is based on a resummation of post-Newtonian results to describe the inspiral, and uses calibration to NR simulations for the late-inspiral and merger~\cite{Zilhao:2013hia,Loffler:2011ay,Lovelace:2010ne,2013PhRvL.111x1104M}. 
Both families describe the ringdown employing results of BH perturbation theory~\cite{Barausse:2011kb,Taracchini:2014zpa}. 
The NR waveforms are obtained by solving the full non-linear Einstein equations and are subject only to numerical errors~\cite{ 2010RvMP...82.3069C}.

Besides the waveform family, models also differ on whether they include the effect of spin-precession~\cite{Apostolatos:1994mx,Kidder:1995zr} and higher-order modes, as indicated in Table~\ref{table:pe-table}. 
From the {\tt IMRPhenom} and {\tt SEOBNR} families we use one aligned-spin and one spin-precessing model without higher-order modes as well as a spin-aligned model with higher-order modes. None of the models from these two waveform families currently include both precessing spins and higher-order modes.
In particular, from the {\tt IMRPhenom} family, we use: the spin-precessing \IMRP{}~\cite{Hannam:2013oca} and the spin-aligned \IMRPD{}~\cite{Husa:2015iqa,Khan:2015jqa} models. 
From the {\tt SEOBNR} family we use: the spin-precessing \SEOBP{}\footnote{\SEOBP{} includes the modes $(\ell, |m|) = (2,2), (2,1)$ in the coprecessing frame, the coordinate system for which the $z$-axis is instantaneously aligned with the Newtonian angular momentum, see~\cite{Buonanno:2002fy}.}~\cite{2014PhRvD..89f1502T,2014PhRvD..89h4006P} and the spin-aligned \SEOBA{}\footnote{In particular, we use the reduced order model implementation of  \SEOBA{} which is computationally less expensive~\cite{Purrer:2014fza,Purrer:2015tud}.}~\cite{Bohe:2016gbl} models. 
As far as higher-order modes are concerned, we use \IMRPHM{}~\cite{London:2017bcn}, a spin-aligned model of the {\tt IMRPhenom} family that includes the $(\ell, |m|) = [(2,2),(2,1),(3,3),(3,2),(4,4),(4,3)]$ higher-order modes and \SEOBHM{}~\cite{Cotesta:2018fcv}, a spin-aligned model of the {\tt SEOBNR} family that includes the $(\ell, |m|) = [(2,1),(3,3),(4,4),(5,5)]$ higher-order modes. 
Posterior samples obtained with \IMRP{} and \SEOBP{} have already been made publicly available by the LIGO-Virgo Collaborations and we use them directly~\cite{GW170729Samples}.

The NR simulations we use include a total of 763 spin-aligned and 625 spin-precessing simulations~\cite{2013PhRvL.111x1104M,2016CQGra..33t4001J,2017CQGra..34v4001H}.
We optionally augment the list of NR simulations with waveforms computed using \NRSur{}~\cite{2015PhRvL.115l1102B,2017PhRvD..95j4023B,2017PhRvD..96b4058B}, a surrogate model directly based on NR.
The surrogate model we use is valid for mass ratio $0.5\le q\le1$ and dimensionless spin magnitude $\chi\le0.8$; however, the analysis performed results in a full posterior due to the inclusion of the NR waveforms that cover the remaining region. For both NR-related analyses, we include results with higher-order modes $(\ell\leq4, |m|\leq\ell)$ and with only the $(\ell=2, |m|\leq\ell)$ modes. 
When using the NR simulations we also assume that the spins remain aligned to the orbital angular momentum (no spin-precession). 

\subsection{Priors} 

Our analysis employs the following priors for the source parameters. 
The detector-frame component masses $m_1, m_2$ are assumed to be uniform between $10M_{\odot}$ and $200M_{\odot}$ with $m_1>m_2$,
while the mass ratio $q\equiv m_2/m_1$ is restricted to be above 0.125.
The sky location and orientation of the binary, as well as the directions of the component spins are uniform on the unit sphere. 
The distance is uniform in volume with a maximum cut off of $7$Gpc, while the time and phase of arrival are uniform. 
We have verified that the mass and distance prior ranges encompass the entire region where the posterior distribution has non-negligible support.

For the magnitude of the dimensionless component spins $\chi_i, \,i\in \{1,2\},$ we make different choices in order to investigate how this affects the posterior. 
The first prior is uniform-in-$\chi$ up to 0.99 for both spin-aligned and spin-precessing waveform models. 
The second prior is uniform-in-$\chi_z$, where $\chi_z$ is the spin projection along the axis perpendicular to the orbital plane, with the restriction that the spin magnitude is below 1. 
For spin-precessing model, the in-plane $\chi_x$ and $\chi_y$ components are also uniform; in that case this prior is sometimes referred to as `volumetric' prior as it corresponds to the spin vector being uniformly distributed within the unit sphere.

Finally, we also use priors targeted toward CBCs where the primary component is the product of a past merger. We study two cases.
The 1g case uses a spin prior that is uniform-in-$\chi$ for both component spins and a mass ratio prior that favors equal masses
\begin{align}
p_{1g}(q) \propto \left\{ 1 + \exp{\left[ -k\left( q-q_0 \right) \right]} \right\}^{-1},
\end{align}
where $k=20$ and $q_0=0.8$. This prior choice was motivated in~\cite{salvoprior}.
In the case of a 2g merger, the primary is expected to be more massive than the secondary binary component. We use a Gaussian mass ratio prior with a mean of $0.5$ and a standard deviation of $0.2$, motivated by Fig. 2 of~\cite{davide2g}.
In the 2g case, the primary is also expected to be spinning more rapidly.
We therefore use a prior where $\chi_1$ is distributed according to a Gaussian centered at $0.7$ with a width of $0.1$~\cite{davide2g,Fishbach:2017dwv}.
The priors of secondary spin magnitude and both spin directions are uniform.

\subsection{{\tt LALInference} and {\tt RIFT}} 

Given a waveform model and a set of prior choices, we compute the joint multidimensional posterior distribution of the source parameters. 
For fast-to-evaluate waveform models we use the publicly available software library {\tt LALInference}~\cite{Veitch:2014wba,lalinference_o2} to directly sample the posterior distribution. 
This approach computes the likelihood exactly at various points of the parameter space, but in order to obtain enough independent samples, millions of likelihood evaluation are required. 
This is prohibitive for models that are slow to evaluate, such as NR. 

In these cases we use {\tt RIFT}~\cite{2015PhRvD..92b3002P,2018arXiv180510457L}. 
{\tt RIFT}'s three-stage algorithm first evaluates the likelihood on a dense grid; then approximates it via interpolation; and then uses Monte Carlo integration to produce the full posterior distribution. 
The number of grid points used for this particular NR-only analysis is 63,000, and the number of added surrogate points for the NR/\NRSur{} grid was
 40,000; this brought the total number of points for the NR/\NRSur{} to 103,000. For context, {\tt RIFT} in general calculates the marginalized likelihood on thousands grid points in parallel. For each marginalized likelihood, which has fixed intrinsic parameters, we evaluate the likelihood at $\approx10^6$ different extrinsic parameters. 
  Even though the number of evaluations are orders of magnitude larger than for {\tt LALInference}, the overall wallclock time is considerably lower because the likelihood evaluations are faster and done in parallel, see \cite{2019arXiv190204934W} for details. However, due to grid limitations (discreteness and limited range), {\tt RIFT} does not sample both extrinsic and intrinsic parameters jointly from the full posterior distribution. Instead it marginalizes over all extrinsic parameters to calculate the likelihood and posterior for just the intrinsic parameters.  

All results obtained using {\tt LALInference} marginalize over the same detector calibration amplitude and phase uncertainty as in~\cite{LIGOScientific:2018mvr} and publicly available in~\cite{GW170729CalEnv} using the method described in~\cite{TheLIGOScientific:2016wfe,SplineCalMarg-T1400682}. 
All {\tt RIFT} results assume perfect calibration; this choice was shown to not affect the intrinsic binary parameters~\cite{Vitale:2011wu}.

\subsection{{\tt BayesWave}} 

Finally, we also use a minimal-assumptions analysis that does not make use of CBC-specific waveform models. We use {\tt BayesWave}~\cite{Cornish:2014kda}, a publicly-available algorithm~\cite{bayeswave} that does not explicitly assume that the signal is a CBC\footnote{While {\tt BayesWave} does not assume an explicit signal morphology, it does assume that the signal is elliptically polarized, that it propagates at the speed of light, and that there is no phase decoherence during the propagation.}, and instead models it through a linear combination of basis functions, either sine Gaussian (known as Morlet Gabor) wavelets, or ``chirplets''~\cite{Millhouse:2018dgi}, as listed in the bottom half of Table~\ref{table:pe-table}. 
The latter are sine Gaussians modified with a linearly evolving frequency. {\tt BayesWave} relies on a transdimensional sampler~\cite{10.1093/biomet/82.4.711} to explore the multidimensional posterior of the parameters of the wavelets/chirplets (frequency, time, phase, amplitude, quality factor, and possibly the frequency derivative) as well as the number of wavelets/chirplets in the linear combination. 

We then compare the signal reconstruction obtained with the morphology-independent models of {\tt BayesWave} and with CBC waveform models. 
Broad agreement between the wavelet reconstruction and {\tt IMRPhenomPv2} was established in~\cite{LIGOScientific:2018mvr} and we here perform the same test for waveform models that include higher-order modes. 
We also quantify the level of consistency through the detector network overlap \cite{Apostolatos:1995pj}, defined as 
\begin{equation}
{\cal{O}}_{N} \equiv \frac{(h_1,h_2)_{N}}{\sqrt{(h_1,h_1)_{N}(h_2,h_2)_{N}}}
\end{equation}
where $(h_1,h_2)_{N}$ denotes the inner product over the network defined by

\begin{equation}
(h_1,h_2)_{N} = \sum_{i}^{n} (h_1^i, h_2^i)
\end{equation}
where $i$ sums over all the detectors in the network, and $(h_1^i,h_2^i)$ is the inner product in an individual detector defined by

\begin{equation}
(h_1^i,h_2^i)\equiv 4 \Re \int_{0}^{\infty} \frac{\tilde{h}_1^i(f)\tilde{h}^{i*}_2(f)}{S_n^i(f)}df.
\end{equation}
In the above, $\tilde{h}_1^i(f)$ denotes a signal reconstruction sample computed with CBC models and $\tilde{h}_2^i(f)$ is a reconstruction sample computed with {\tt BayesWave}. Finally, $S_n^i(f)$ denotes the PSD of the detector. The superscipt $i$ denotes the quantities as they appear in the the $i^{\rm{th}}$ detector.

\section{CBC-model-based analysis}
\label{CBCresults}

In this section we present results on inference with CBC waveform models. 
We study how the posteriors for the various source parameters are affected by the inclusion of spin-precession and higher-order modes in the waveforms. 
We also study the effect of spin priors and waveform systematics on the validity of our conclusions. 

\subsection{Higher-order modes}
\label{2HM}

The importance of higher-order modes on a GW signal observed in the detectors depends on both extrinsic parameters, such as the inclination of the binary, and intrinsic parameters, such as the mass ratio. 
In general, signals from edge-on and asymmetric binaries include more power in higher-order modes. 
To study the effect of higher-order modes on GW170729 we analyze the data with waveform models both with and without higher-order modes. 
Table~\ref{table:par-table} gives the median and 90\% symmetric credible interval and/or HPD (highest probability density interval) for various source parameters from multiple waveform models  and the two spin priors.

\begin{table*}[]
\centering
  \begin{tabular}{cccccccccccccc}
    \hline \hline
Parameter   &$m_{1} (M_{\odot})$ & $m_{2} (M_{\odot})$ & $M (M_{\odot})$ &$q$ &$\chi_{\rm{eff}}$ & $\chi_{\rm{p}}$ & SNR & $D_L(\text{Gpc})$ & $\vert \cos{\theta_{JN}}\vert$\\ \hline
{\tt IMRPhenomPv2 ($\chi$)} & $51.0^{+13.9}_{-12.4}$& $31.9^{+9.3}_{-9.6}$& $83.7^{+13.0}_{-12.0}$& $0.62^{+0.36}_{-0.23}$& $0.35^{+0.22}_{-0.23}$& $0.42^{+0.34}_{-0.29}$& $10.7^{+0.4}_{-0.4}$& $2.85^{+1.31}_{-1.28}$& $0.83^{+0.17}_{-0.40}$ \\
{\tt IMRPhenomPv2 ($\chi_z$)} & $52.0^{+15.9}_{-11.6}$& $33.2^{+9.9}_{-9.7}$& $85.8^{+12.4}_{-12.4}$& $0.64^{+0.34}_{-0.25}$& $0.41^{+0.21}_{-0.21}$& $0.58^{+0.29}_{-0.29}$& $10.7^{+0.4}_{-0.4}$& $2.96^{+1.30}_{-1.38}$& $0.84^{+0.16}_{-0.42}$ \\
{\tt IMRPhenomD ($\chi$)} & $50.5^{+13.5}_{-11.2}$& $32.5^{+10.0}_{-8.8}$& $82.8^{+13.4}_{-12.6}$& $0.65^{+0.32}_{-0.23}$& $0.34^{+0.21}_{-0.23}$& -& $10.7^{+0.4}_{-0.4}$& $2.75^{+1.26}_{-1.41}$& $0.80^{+0.20}_{-0.46}$ \\
{\tt IMRPhenomD ($\chi_z$)} & $50.8^{+12.8}_{-11.5}$& $34.5^{+9.7}_{-9.2}$& $85.5^{+13.9}_{-11.4}$& $0.68^{+0.29}_{-0.25}$& $0.42^{+0.19}_{-0.21}$& -& $10.8^{+0.4}_{-0.3}$& $2.90^{+1.32}_{-1.48}$& $0.79^{+0.21}_{-0.46}$ \\
{\tt IMRPhenomHM} ($\chi$) & $57.0^{+12.6}_{-12.0}$& $29.5^{+9.2}_{-9.0}$& $86.0^{+12.2}_{-12.0}$& $0.52^{+0.26}_{-0.21}$& $0.27^{+0.23}_{-0.27}$& -& $11.1^{+0.4}_{-0.4}$& $2.15^{+1.19}_{-1.15}$& $0.70^{+0.30}_{-0.42}$ \\
{\tt IMRPhenomHM}  ($\chi_z$) & $58.1^{+12.5}_{-13.4}$& $32.0^{+9.5}_{-8.9}$& $89.7^{+13.4}_{-12.9}$& $0.55^{+0.26}_{-0.21}$& $0.36^{+0.21}_{-0.23}$& -& $11.1^{+0.4}_{-0.4}$& $2.30^{+1.24}_{-1.22}$& $0.68^{+0.32}_{-0.46}$ \\
{\tt SEOBNRv3 ($\chi$)} & $49.5^{+13.2}_{-10.8}$& $35.3^{+8.9}_{-8.6}$& $85.0^{+13.8}_{-12.8}$& $0.72^{+0.28}_{-0.22}$& $0.38^{+0.23}_{-0.22}$& $0.45^{+0.32}_{-0.31}$& $10.8^{+0.4}_{-0.4}$& $2.86^{+1.38}_{-1.38}$& $0.78^{+0.22}_{-0.47}$ \\
{\tt SEOBNRv4 ($\chi$)} & $50.8^{+12.4}_{-11.8}$& $33.4^{+9.6}_{-9.8}$& $83.9^{+13.7}_{-13.3}$& $0.66^{+0.30}_{-0.25}$& $0.35^{+0.20}_{-0.24}$& -& $10.8^{+0.4}_{-0.3}$& $2.77^{+1.30}_{-1.45}$& $0.79^{+0.21}_{-0.46}$ \\
{\tt SEOBNRv4 ($\chi_z$)} & $51.7^{+12.4}_{-12.6}$& $35.2^{+9.4}_{-9.6}$& $86.4^{+14.9}_{-12.7}$& $0.68^{+0.29}_{-0.24}$& $0.42^{+0.21}_{-0.26}$& -& $10.8^{+0.4}_{-0.3}$& $2.97^{+1.46}_{-1.43}$& $0.79^{+0.21}_{-0.47}$ \\
{\tt SEOBNRv4HM}  ($\chi$) & $55.2^{+10.2}_{-12.2}$& $29.8^{+9.9}_{-9.5}$& $84.6^{+12.5}_{-11.3}$& $0.54^{+0.31}_{-0.20}$& $0.25^{+0.26}_{-0.26}$& - & $11.0^{+0.4}_{-0.4}$& $2.30^{+1.36}_{-1.17}$ & $0.72^{+0.28}_{-0.39}$ \\
{\tt SEOBNRv4HM}  ($\chi_z$) & $54.8^{+11.1}_{-12.1}$& $32.8^{+9.8}_{-9.2}$& $87.5^{+12.2}_{-11.7}$& $0.60^{+0.32}_{-0.21}$& $0.34^{+0.26}_{-0.24}$& - & $11.0^{+0.4}_{-0.4}$& $2.66^{+1.38}_{-1.31}$ & $0.74^{+0.26}_{-0.41}$ \\
\end{tabular}
\caption{Parameters of GW170729 obtained with various waveform models and two spin priors, uniform-in-$\chi$ (labelled $\chi$) and uniform-in-$\chi_z$ (labelled $\chi_z$). 
We quote median values and 90\% credible intervals for the source-frame primary mass, the source-frame secondary mass, the source-frame total mass, the effective spin $\chi_{\rm{eff}}$, and the effective precession parameter $\chi_{\rm{p}}$~\cite{Schmidt:2014iyl}. 
For the mass ratio we quote the median value and the 90\% HPD. 
All masses are given in the source frame assuming the cosmological parameters of~\cite{Ade:2015xua} to convert luminosity distance to redshift.
The effective precession parameter $\chi_{\rm{p}}$ is absent in the spin-aligned models.
}
\label{table:par-table}
\end{table*}

\begin{figure*}[]
\includegraphics[width=\columnwidth,clip=true]{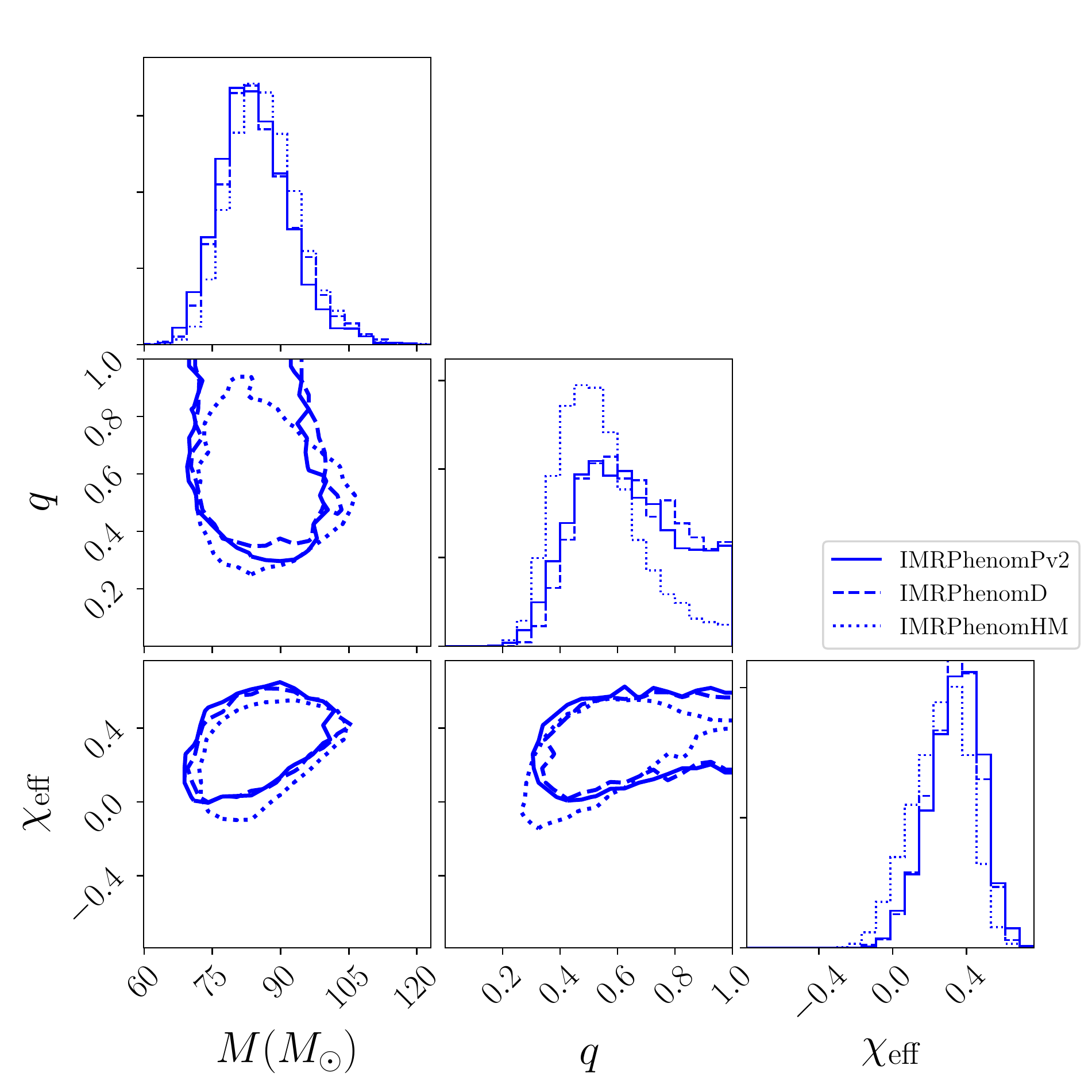}
\includegraphics[width=\columnwidth,clip=true]{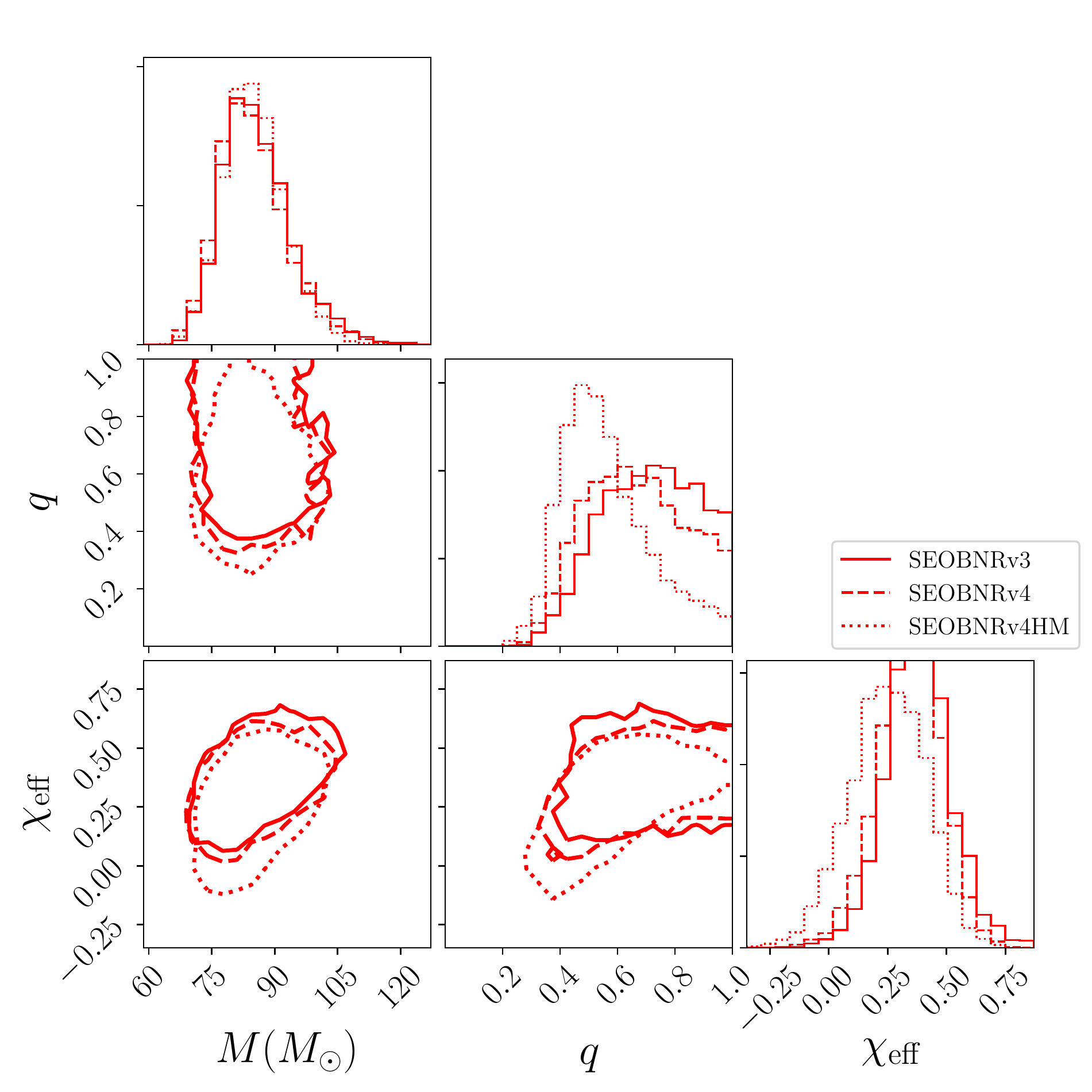}
\caption{Corner plots for the posterior densities of the total binary mass in the source frame, the mass ratio, and the effective spin parameter for the different waveform families. 
We show results obtained with CBC waveforms of the {\tt IMRPhenom} family (left) and the {\tt SEOBNR} family (right).
Results from both waveform families suggest that the inclusion of higher-order modes in the models results in evidence for more unequal-mass binaries and less support for nonzero effective spin. 
The total mass of the binary remains unchanged.}
\label{fig:intrinsic_IMREOB}
\end{figure*}

We start by discussing the binary intrinsic parameters, in particular masses and spins. 
In Fig.~\ref{fig:intrinsic_IMREOB} we show multi-dimensional corner plots for the posterior densities of the source-frame total mass $M$, the mass ratio $q$ and the effective spin parameter $\chi_{\rm{eff}}$ with waveform models of the {\tt IMRPhenom} (left) and the {\tt SEOBNR} family (right). 
For these figures and all the figures of this subsection we show results with the uniform-in-$\chi$ spin prior. 
Within each family, the spin-precessing model is given with a solid line, the spin-aligned model with a dashed line and the spin-aligned model with higher-order modes with a dotted line. 

In all cases we find that the inclusion of higher-order modes does not have a large effect on the total mass measurement. 
The mass ratio of the system and the effective spin posteriors are both shifted. 
In particular we find that waveforms with higher-order modes consistently provide more support for unequal-mass systems and smaller effective spins. 
For the mass ratio we find that the 90\% HPD interval is $(0.31-0.78) [(0.34-0.85)]$ with higher-order modes and $(0.42-0.97) [(0.41-0.96)]$ without them when using the spin-aligned {\tt IMRPhenomHM} [{\tt SEOBNRv4HM}] and {\tt IMRPhenomD} [{\tt SEOBNRv4}] models respectively. We conclude that GW170729 is not consistent with an equal-mass merger at the 90\% level at least. 

The combination of unchanged total mass but lower mass ratio means that the primary mass of GW170729 is inferred to be larger than previously measured. 
Reference~\cite{LIGOScientific:2018jsj} studied the population of the 10 detected BBHs and concluded that no more than 1\% of BHs in BBHs are expected to be above $45 M_{\odot}$.
We find that our updated primary mass measurement is not at odds with this conclusion. 
In particular, we find that the probability that $m_1$ is lower than $45 M_{\odot}$ is 17\% using {\tt IMRPhenomPv2} and reduced to 6\% with {\tt IMRPhenomHM}.
Given that we have detected 10 BBHs, it is not unlikely that the true $m_1$ of one of them is at the sixth posterior percentile.
A more detailed population analysis in needed to quantify this statement, but this is beyond the scope of this paper.

We also find that higher-order modes result in less support for a positive effective spin in GW170729. 
Reference~\cite{LIGOScientific:2018mvr} reported that for GW170729 $\chi_{\rm{eff}} \sim (0.11- 0.58)$ at the 90\% credible level using combined posterior samples between {\tt IMRPhenomPv2} and  {\tt SEOBNRv3}. 
Interestingly, this credible interval does not include zero, suggesting that at the 90\% level GW170729 has a nonzero effective spin.
The inclusion of higher-order modes slightly changes this picture as we now find that the corresponding 90\% credible intervals no longer exclude zero: $\chi_{\rm{eff}} \sim(-0.01-0.49)$ with \IMRPHM{} and $\chi_{\rm{eff}} \sim(-0.02-0.50)$ with \SEOBHM{}. 
The effective spin parameter is still probably positive with the probability of $\chi_{\rm{eff}}>0$ being $94\%$ with higher-order modes, which is slightly reduced from the corresponding probability of $99\%$ when higher-order modes are not taken into account.
Overall higher-order modes cause the 95\% lower limit for $\chi_{\rm{eff}}$ to shift by $-0.10$ for the {\tt IMRPhenom} family,  $-0.13$ for the {\tt SEOBNR} family.

\begin{figure*}[]
\includegraphics[width=\columnwidth,clip=true]{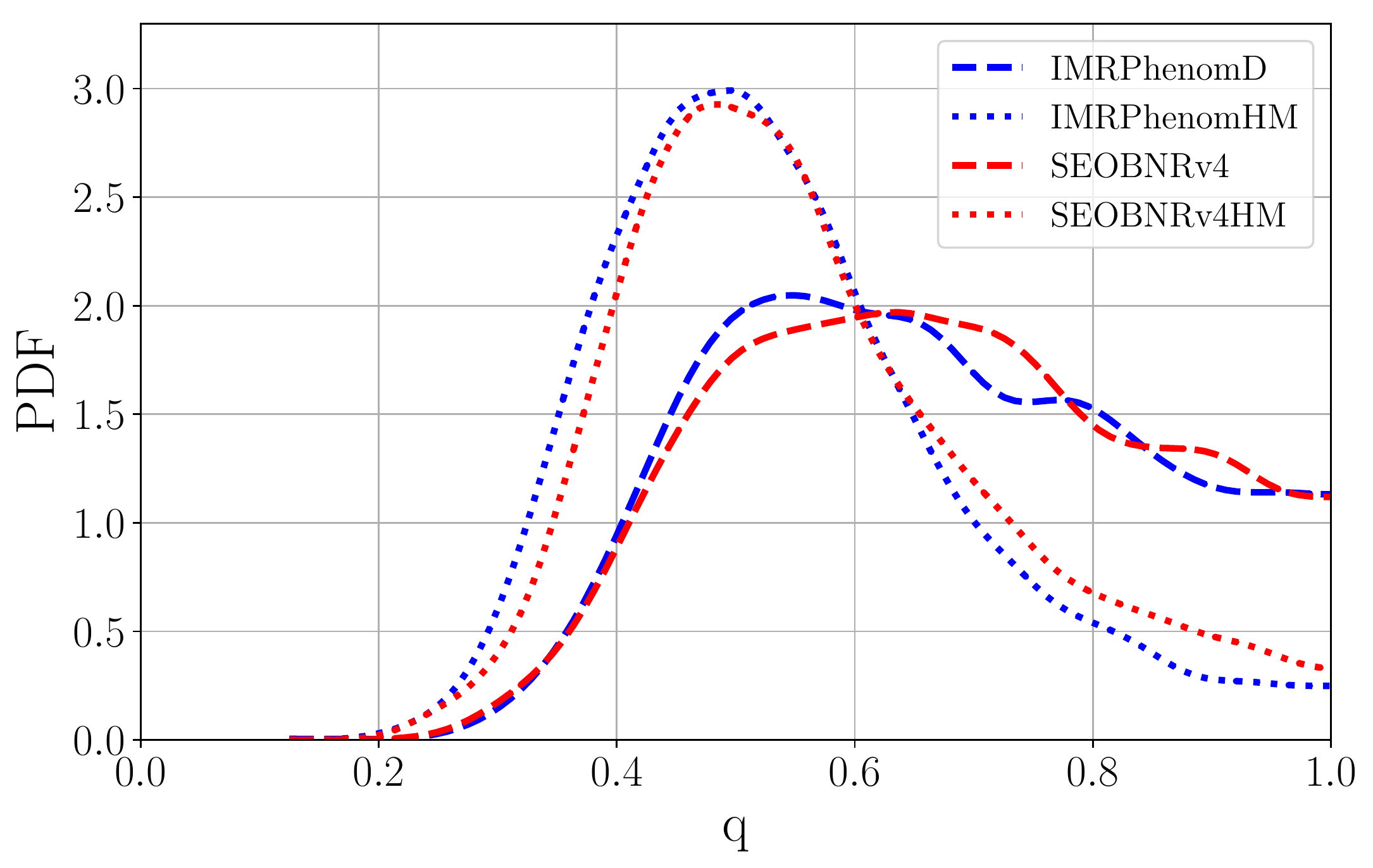}
\includegraphics[width=\columnwidth,clip=true]{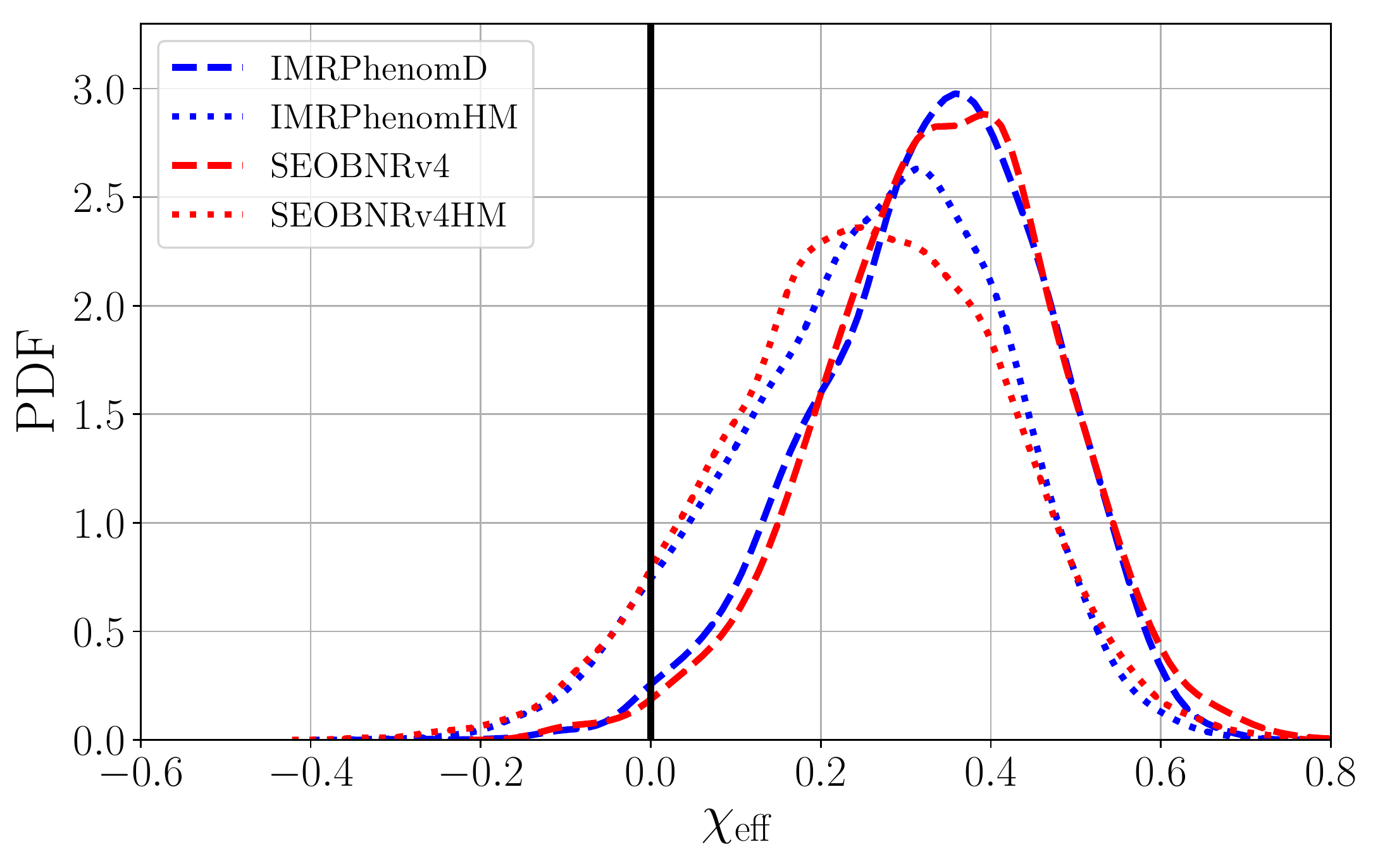}
\caption{Effect of waveform systematics. 
We show the mass ratio posterior (left) and the effective spin posterior (right) computed with different CBC waveform models that include higher-order modes (dotted lines) and models that do not include higher-order modes (dashed). 
Small differences between the posteriors from different waveform approximants are present, but these differences are much smaller than the effect of higher-order modes for both parameters. }
\label{fig:systematics}
\end{figure*}

Despite the broad consistency between results obtained with different waveform families, it is still possible that our results are partly affected by systematic uncertainties in the waveform models.
To address this in Fig.~\ref{fig:systematics} we plot again the mass ratio (left) and effective spin (right) posterior densities for the 
{\tt IMRPhenom} and {\tt SEOBNR} waveform models with (dashed lines) and without (dotted lines) higher-order modes. 
We find very small differences between both all the dotted and all the dashed lines and in particular between the new waveform
models {\tt IMRPhenomHM} and {\tt SEOBNRv4HM} that include higher-order modes.
More importantly, we find a clear separation between the dotted lines, i.e. the posteriors that include higher-order modes, and the dashed lines, i.e. the posteriors that do not include higher-order modes. 

\begin{figure}[]
\includegraphics[width=\columnwidth,clip=true]{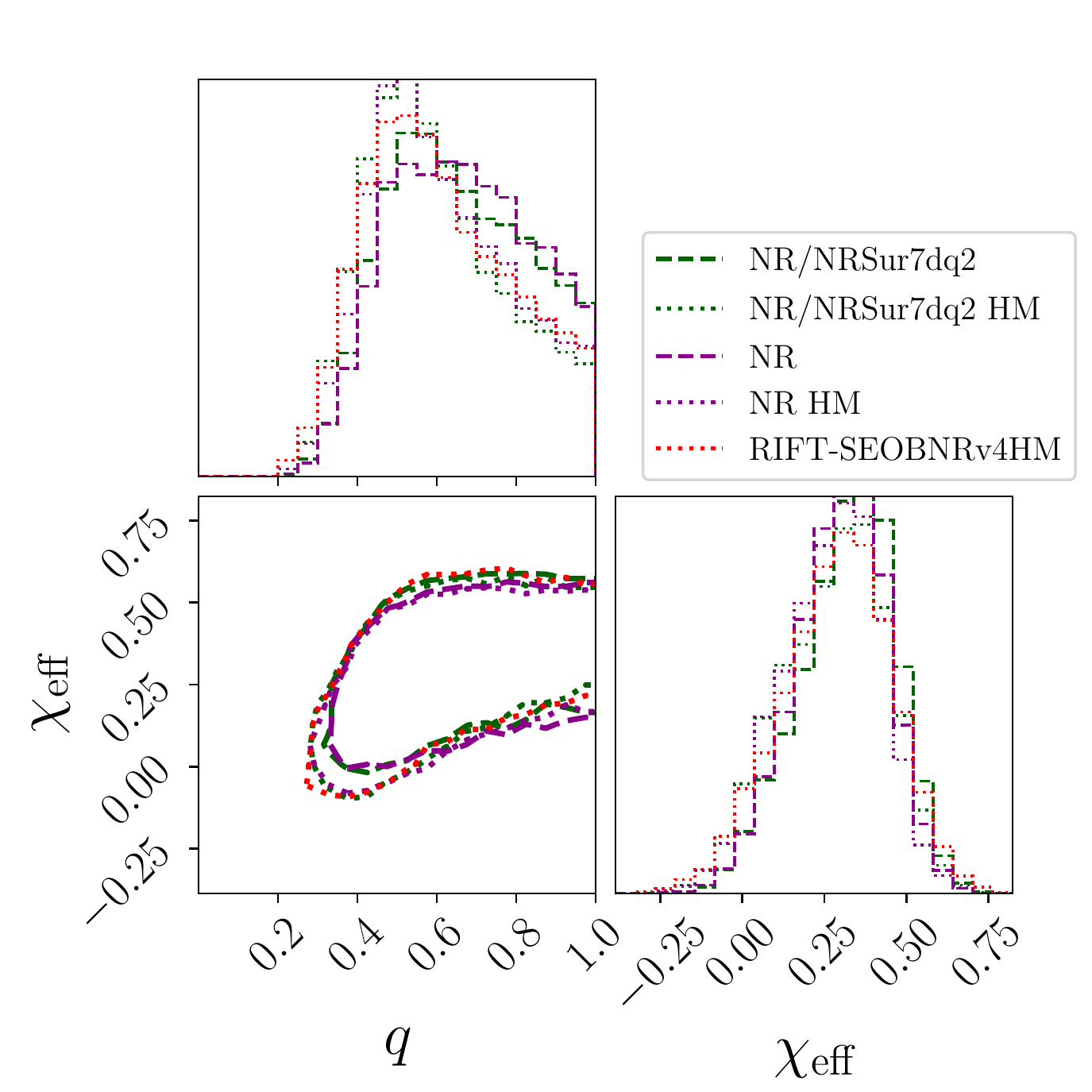}
\caption{Corner plots for the posterior densities of the mass ratio and the effective spin parameter for the NR waveform family with (magenta lines) and without (green lines) \NRSur{} and {\tt SEOBNRv4HM} (red line) computed with {\tt RIFT}. 
As before, we use dashed (dotted) lines for posteriors with (without) higher order modes. 
We observe excellent agreement between NR and NR/\NRSur{}, confirming the high accuracy of the NR surrogate model.
Additionally, we find very good agreement between the NR analysis and {\tt SEOBNRv4HM} when both are used with the
same inference code, {\tt RIFT}. 
}
\label{fig:intrinsic_NR}
\end{figure}

As a further test, in Fig.~\ref{fig:intrinsic_NR} we compare posteriors for  $q$ and $\chi_{\rm{eff}}$ computed 
with {\tt SEOBNRv4HM} (red line)  with NR waveforms (magenta lines), as well as with NR/\NRSur{} (green lines).
Compared to Fig.~\ref{fig:intrinsic_IMREOB} we omit the total mass posterior as {\tt RIFT} did not compute source-frame quantities.
In order to perform a fair comparison, all posteriors have been computed with {\tt RIFT}, while for technical reasons we cannot
use {\tt IMRPhenomHM} with {\tt RIFT}. We find excellent agreement between NR with higher-order modes, NR/\NRSur{} with higher-order modes, and {\tt SEOBNRv4HM}. This shows that {\tt SEOBNRv4HM} is as accurate as NR waveforms in 
describing GW170729. Moreover, the agreement between {\tt SEOBNRv4HM} and {\tt IMRPhenomHM} in Fig.~\ref{fig:systematics}
suggests that the latter {\tt IMRPhenom} waveform is also highly accurate for the event studied here.
While these posteriors are broadly consistent with those obtained in Fig.~\ref{fig:systematics}, we find a disagreement in results obtained 
with {\tt LALInference} and {\tt RIFT} with the same waveform model for the mass ratio at the $7\%$ level. 
The nature of this difference and further numerical estimates are described in Appendix~\ref{RIFTtroubles}.
Overall, Figures~\ref{fig:systematics} and~\ref{fig:intrinsic_NR} suggest that despite minor differences between the waveform models considered here, our main conclusions are robust.

\begin{figure}[]
\includegraphics[width=\columnwidth,clip=true]{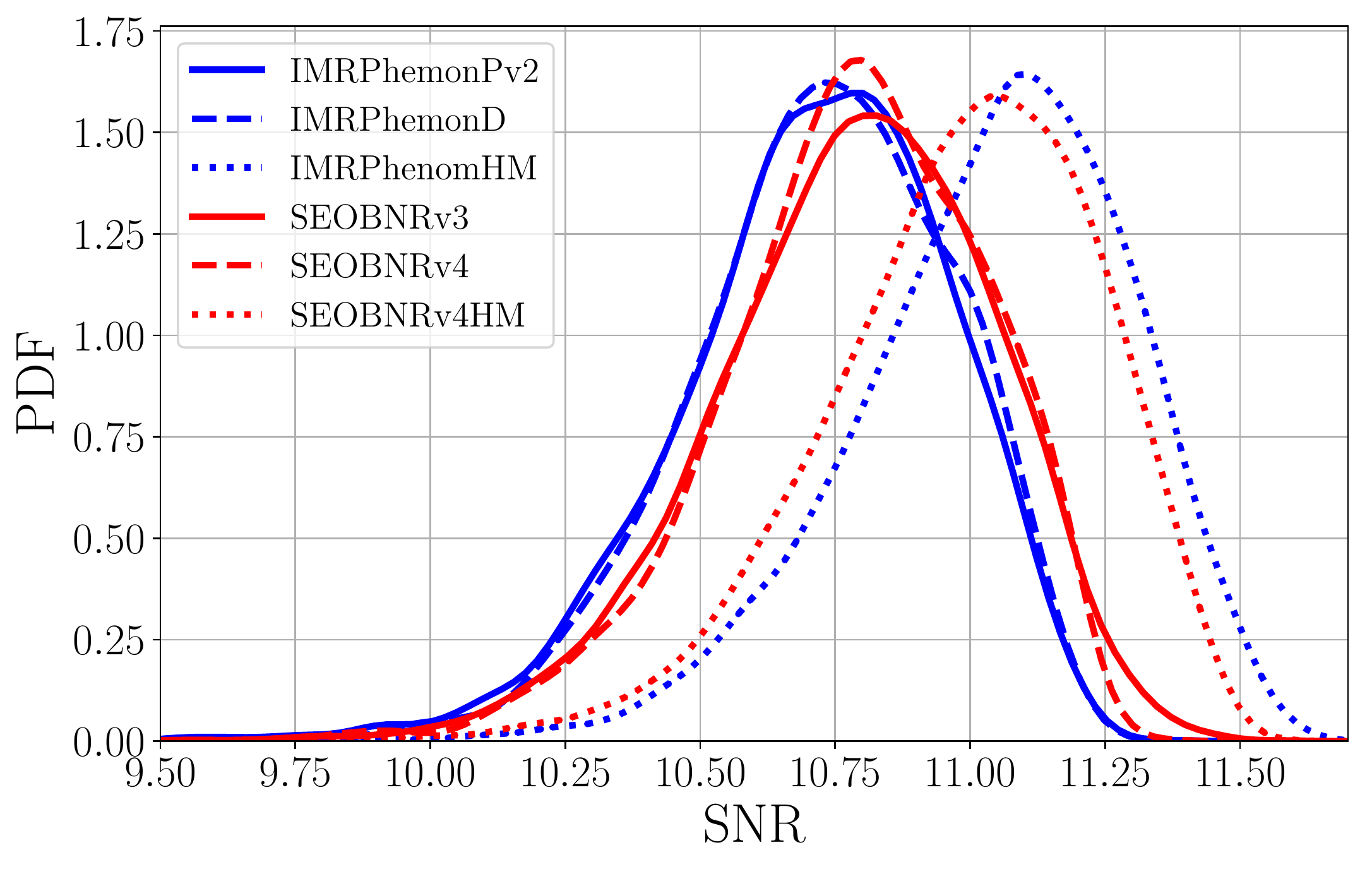}\\
\includegraphics[width=\columnwidth,clip=true]{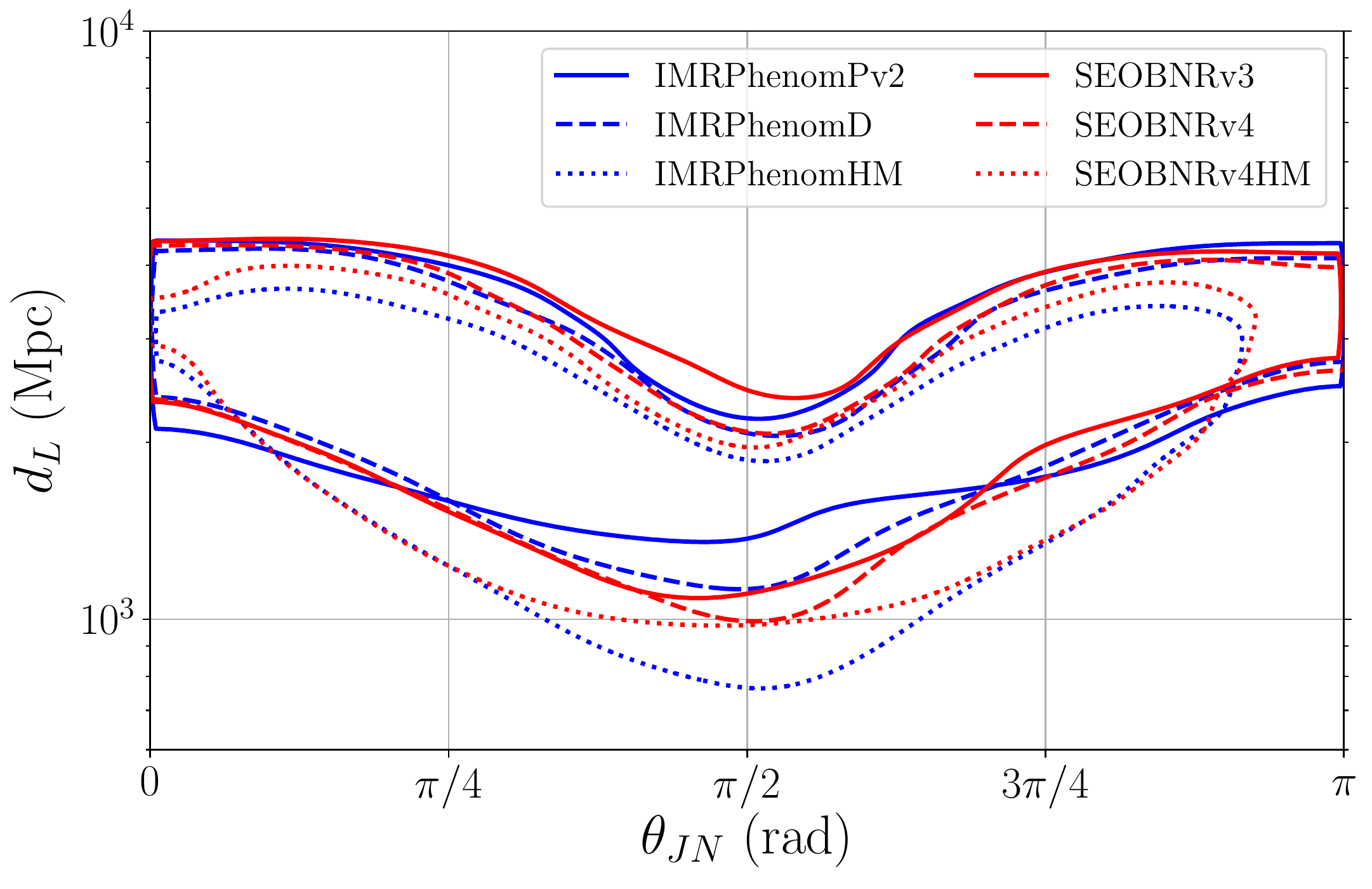}
\caption{Posterior density for the matched-filter SNR (top) and distance-inclination (bottom) for various waveform models. 
We find that both within the {\tt IMRPhenom} and the {\tt SEOBNR} waveform families, the waveform model that includes higher-order modes returns the highest value of matched-filter SNR. 
This higher SNR translates to lower distance and inclination angles closer to 90 degrees.
}
\label{fig:extrinsic}
\end{figure}

We now turn to the binary extrinsic parameters. 
Figure~\ref{fig:extrinsic} shows the matched-filter network signal-to-noise ratio (SNR) posterior density on the top and the two-dimensional posterior density for the luminosity distance and the inclination on the bottom. 
The network SNR is defined as the square root of the squared sums of the matched-filter SNR in each interferometer, calculated as $\rho = (h,d)/\sqrt{(h,h)}$, where $d$ is the data and $h$ the signal model.
The inclination is defined as the angle between the \emph{total} angular momentum vector of the binary, whose direction we treat as fixed,  and the line of sight. 
We present results for waveform models of the {\tt IMRPhenom} (blue) and the {\tt SEOBNR} (red) families. 

The SNR depends on both the intrinsic loudness of the data and the agreement between the signal and the template $h$. 
Since the data are common for all analyses, a larger SNR indicates a better agreement/overlap between the data and the template. 
While we find that within both families spin-precession has a minimal impact on the SNR, waveforms with higher-order modes report slightly larger SNR values, see the top panel of Fig.~\ref{fig:extrinsic}. 
This suggests that their inclusion leads to a marginally better fit of the data. 
The data, though, include both the GW170729 signal and a random realization of Gaussian noise, so a better fit of the data does not necessarily imply that waveforms with higher-order modes recover a larger fraction of the GW signal.

We quantify the impact of higher-order modes on GW170729 by computing the Bayes Factor in favor of {\tt IMRPhenomHM} compared to {\tt IMRPhenomD}. 
We find a BF of 5.1:1. While it favors the model with HM, this BF is consistent with the fact that the HM waveforms are able to extract marginally more SNR from the data\footnote{The BF is related to the SNR through $\log \rm{BF} \propto 1/2\, \rm{SNR}^2$. Therefore an SNR increase of $\sim 0.2$ compared to $\sim 10.8$ (see Table~\ref{table:par-table}) would result in a BF of $\sim 8$. This suggests that the measured BF of $5$ is consistent with the SNR increase due to HM.}.
Moreover, we emphasize that the BF is not the same as the odds ratio in favor of higher-order modes, which quantifies our degree of belief that higher-order modes are present in the signal. 
The odds ratio is the BF times the prior odds in favor of the presence of higher-order modes. 
The latter is formally infinite within GR, as the theory of gravity unequivocally predicts that higher-order modes are present in all CBC signals. 
The BF presented here only quantifies if higher-order modes are a necessary feature of the models in order to describe the data, and not whether we believe that they exist in general.

Regarding the bottom panel of Fig.~\ref{fig:extrinsic}, we find that waveforms with higher-order modes result in less support for face-on/off binary orientations. 
This observation, coupled to the fact that we see more support for unequal masses and lower spins, see Fig.~\ref{fig:intrinsic_IMREOB}, suggests that higher-order modes lead to more support for sources that are intrinsically of lower amplitude\footnote{We have also verified this by computing the posterior of the intrinsic loudness (defined as the product of the SNR and the distance) with and without higher-order modes. 
We find that higher-order modes lead to larger probability for intrinsically quieter sources than the  quadrupole templates.}. 
This in turn leads to a posterior distribution for the luminosity distance that is shifted to lower values, as compared to analyses without higher-order modes, as also seen in the bottom panel of Fig~\ref{fig:extrinsic}.

Overall, we find that the inclusion of higher-order modes induces small but noticeable shifts in the parameter posteriors. 
Specifically, the matched-filter SNR increases, the mass ratio posterior obtains more support for unequal masses, and the effective spin parameter is more consistent with lower values; parameter measurements are given in Table~\ref{table:par-table}. 
The general consistency between the waveform model families we study here shows that our conclusions are robust against waveform systematics.

\subsection{Spin prior}

\begin{figure*}[]
\includegraphics[width=\columnwidth,clip=true]{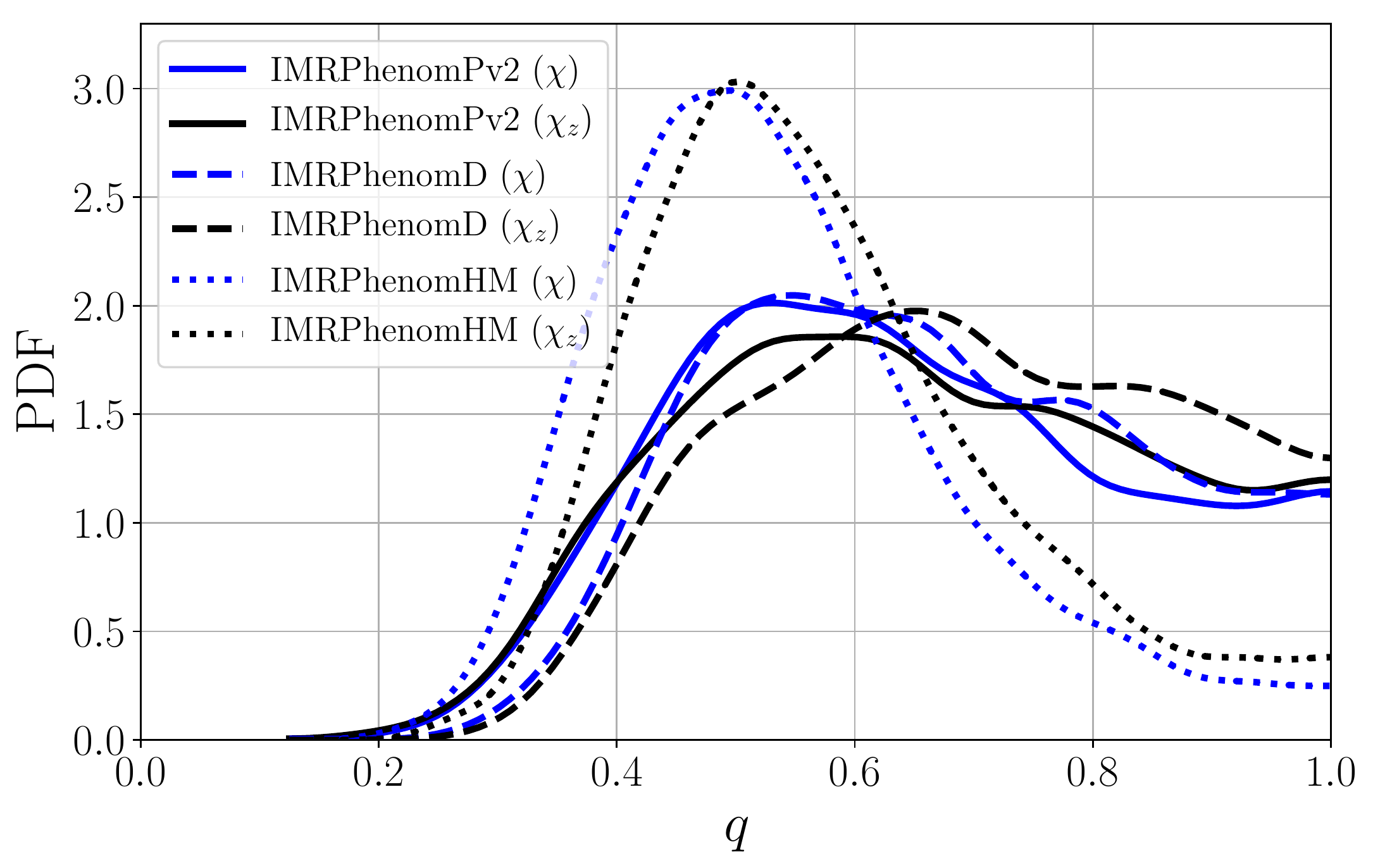}
\includegraphics[width=\columnwidth,clip=true]{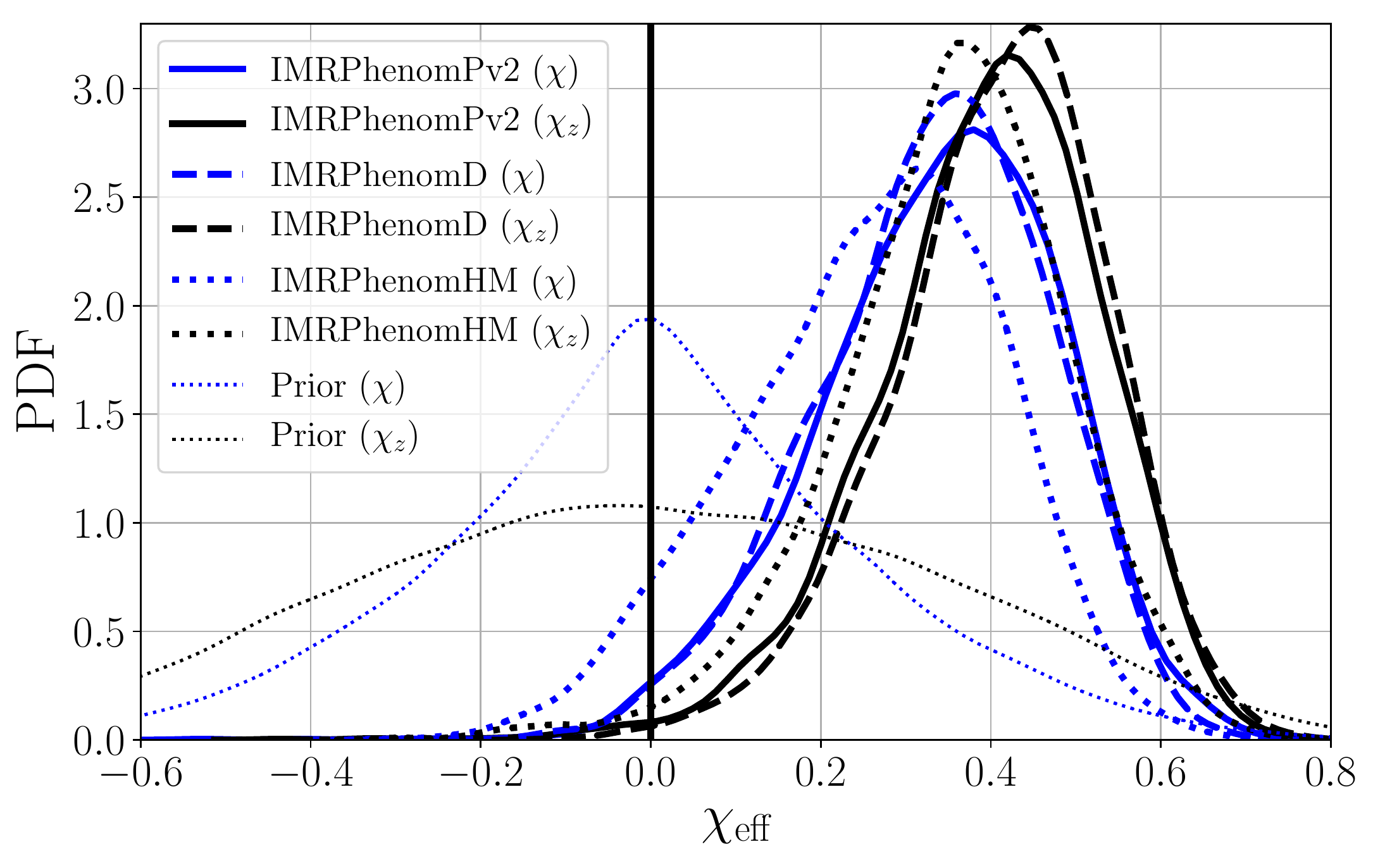}\\
\caption{Effect of spin prior on the mass ratio and effective spin of GW170729. 
As expected, we find that the mass ratio is minimally affected by the spin prior. 
The effective spin, however, is shifted to larger values with the uniform-in-$\chi_z$ prior, resulting in increased evidence for nonzero spins. 
We show results obtained with {\tt IMRPhenom} waveform models, but obtain similar results with {\tt SEOBNR} and NR waveform models as well.}
\label{fig:prior}
\end{figure*}

Besides the CBC waveform models, posterior measurements are also affected by prior choices, in particular the spin prior~\cite{salvoprior}. 
To test the effect of the spin prior, we reanalyze the data this time assuming a uniform-in-$\chi_z$ prior, where $\chi_z$ is the spin component perpendicular to the orbital plane. 
The results are presented in Fig.~\ref{fig:prior} for the mass ratio (left) and the effective spin (right) and for waveforms of the {\tt IMRPhenom} family. 
We have verified that we obtain qualitatively similar results when using {\tt SEOBNR} and NR waveforms models. 
Due to computational constraints we have only checked results with {\tt SEOBNRv3} and the uniform-in-$\chi$ prior, as computed in~\cite{LIGOScientific:2018mvr}.

We find that the spin prior has a minimal effect on the mass ratio posterior. 
This is expected as the correlation between mass ratio and effective spin is mostly present in the inspiral phase of a CBC. 
High-mass systems, such as GW170729 are instead dominated by the merger and ringdown in the LIGO sensitive frequency band. 
In this case little correlation exists between mass ratio and effective spin~\cite{Ng:2018neg}, and changing the spin prior doesnt affect the mass ratio posterior.
The effective spin parameter, on the contrary, is directly affected by the choice of the spin prior, and clear differences are visible. 
The uniform-in-$\chi_z$ prior favors larger spin magnitudes than the uniform-in-$\chi$ prior. 
As a result, the effective spin posterior is shifted to larger values. 
The median and 90\% credible interval for the effective spin is $0.41^{+0.21}_{-0.21}$ under the uniform-in-$\chi_z$ prior and $0.35^{+0.22}_{-0.23}$ under the uniform-in-$\chi$ prior using the {\tt IMRPhenomPv2} waveform model. 
Additional spin measurements for other waveform models are presented in Table~\ref{table:par-table}.

\subsection{Second generation merger}

\begin{figure*}[]
\includegraphics[width=\columnwidth,clip=true]{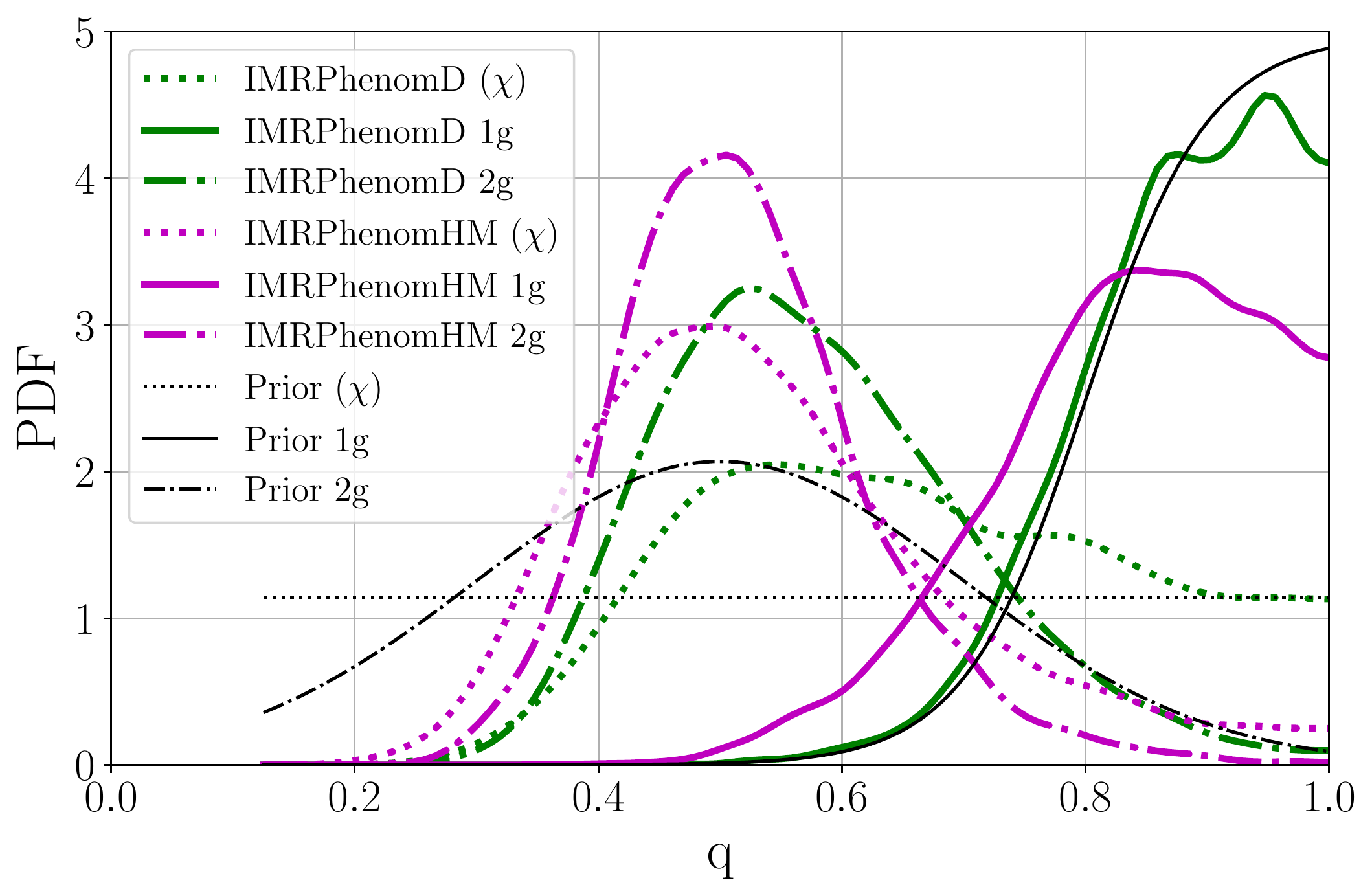}
\includegraphics[width=\columnwidth,clip=true]{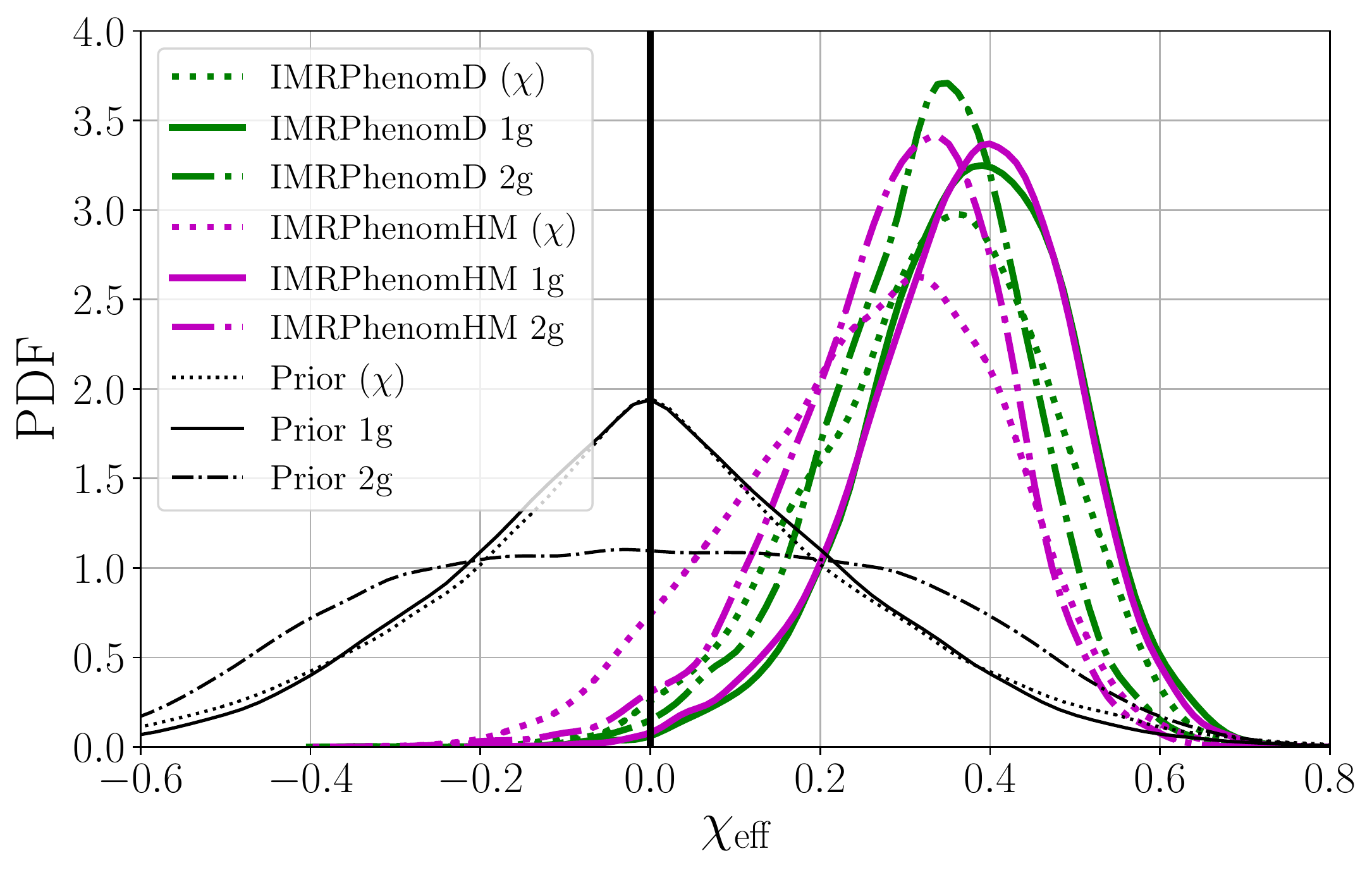}
\caption{Posterior density for the mass ratio (left) and the effective spin (right) under the 1g prior (solid lines) and the 2g prior (dashed lines) for various waveforms. 
The black lines show the prior density distributions for each parameter. 
We also show the uniform-in-$\chi$ posterior and prior from Fig.~\ref{fig:prior} for comparison.
We find that the 1g-2g prior has a minimal effect on the effective spin parameter, but affects the mass ratio considerably. 
All 2g runs have strong support for unequal masses, as expected from the prior.}
\label{fig:2G}
\end{figure*}

Finally, we study the possibility of a 2g merger. 
In that scenario, GW170729 is created in a dense environment such as a nuclear or a globular cluster and its primary mass is the product of a previous merger of two BHs~\cite{davide2g,Fishbach:2017dwv,DiCarlo:2019pmf}. 
In that case, the mass ratio of the system is expected to be closer to 2:1 (as we find when we use waveform models with HM) and the spin magnitude of the primary should be close to 0.7 (the typical spin of the remnant BH after the merger of two equal-mass, 
nonspinning BHs)~\cite{Gonzalez:2006md,Berti:2007fi,Buonanno:2007sv,Campanelli:2006gf,Baker:2003ds}
\footnote{Though a 2g merger scenario provides a simple way to produce a highly spinning BH, 
such systems could also be produced in certain astrophysical scenarios, e.g~\cite{Bouffanais:2019nrw}.}.

We repeat our analysis with two more priors tailored to the cases of a 1g and a 2g merger scenario. 
Table~\ref{table:par-table-2g} gives the median and 90\% symmetric credible interval and/or HPD for various source parameters obtained with the 1g and 2g priors.
Figure~\ref{fig:2G} shows the effect of this 2g prior on the mass ratio (left) and the effective spin (right) of the binary when using {\tt IMRPhenomD} and {\tt IMRPhenomHM}.  
We have verified that we obtain similar results with {\tt SEOBNR} and NR waveform models.
As expected from the priors we have selected, 2g runs show strong support for unequal masses. 
This support is even more evident for waveforms with higher-order modes, as anticipated from the results of the previous section. 
The effective spin parameter is similar, with 1g runs showing more support for nonzero binary components.

To further quantify the prior effect, we calculate the Kullback-Leibler (KL) divergence of posterior against prior~\cite{kullback1951}.
We find a KL divergence for the mass ratio in the 1g case that is $\sim10$ times smaller than the KL divergence in 2g or default prior scenarios, implying that we do not extract much information by applying the 1g mass ratio prior.
For the effective spin, we find that the KL divergence in both priors are comparable as expected because the prior on the effective spin does not change as drastically as the prior on the mass ratio.

We also compute the 2g-vs-1g BF.
We find 4.7:1 (1.4:1) for waveforms with (without) higher-order modes. We conclude that there is not enough support for the hypothesis that the GW170729 primary needs to be the product of a past merger in order to explain the data's properties.
Note that both 2g and 1g models have the same number of parameters, but different distributions in mass ratio and primary spin. The resulting BFs are therefore only affected by how well each model fits the data.

\begin{table*}[]
\centering
  \begin{tabular}{cccccccccccccc}
    \hline \hline
Parameter   &$m_{1} (M_{\odot})$ & $m_{2} (M_{\odot})$ & $M (M_{\odot})$ &$q$ &$\chi_{\rm{eff}}$ & $\chi_{\rm{p}}$ & SNR & $D_L(\text{Gpc})$ & $\vert \cos{\theta_{JN}}\vert$\\ \hline
{\tt IMRPhenomPv2 (1g)} & $43.9^{+8.3}_{-6.7}$& $38.0^{+7.0}_{-6.3}$& $82.1^{+12.9}_{-11.6}$& $0.88^{+0.12}_{-0.14}$& $0.39^{+0.20}_{-0.22}$& $0.47^{+0.31}_{-0.31}$& $10.7^{+0.4}_{-0.4}$& $3.08^{+1.17}_{-1.47}$& $0.83^{+0.17}_{-0.42}$\\ 
{\tt IMRPhenomPv2 (2g)} & $55.1^{+11.3}_{-12.6}$& $30.2^{+8.0}_{-7.2}$& $85.5^{+12.1}_{-11.2}$& $0.55^{+0.23}_{-0.20}$& $0.37^{+0.18}_{-0.19}$& $0.48^{+0.25}_{-0.29}$& $10.7^{+0.4}_{-0.4}$& $2.83^{+1.19}_{-1.31}$& $0.83^{+0.17}_{-0.40}$\\
{\tt IMRPhenomD (1g)} & $43.8^{+7.9}_{-7.2}$& $38.2^{+6.9}_{-6.4}$& $81.9^{+13.9}_{-11.2}$& $0.88^{+0.12}_{-0.13}$& $0.38^{+0.18}_{-0.20}$& -& $10.8^{+0.4}_{-0.4}$& $2.99^{+1.26}_{-1.43}$& $0.80^{+0.20}_{-0.44}$\\
{\tt IMRPhenomD (2g)} & $53.7^{+11.0}_{-11.6}$& $30.2^{+7.9}_{-7.0}$& $84.0^{+13.3}_{-11.8}$& $0.56^{+0.21}_{-0.19}$& $0.34^{+0.17}_{-0.20}$& -& $10.8^{+0.4}_{-0.3}$& $2.69^{+1.23}_{-1.31}$& $0.79^{+0.21}_{-0.44}$\\
{\tt IMRPhenomHM (1g)} & $45.5^{+10.0}_{-7.9}$& $37.9^{+6.7}_{-6.4}$& $83.6^{+13.3}_{-11.7}$& $0.84^{+0.16}_{-0.16}$& $0.38^{+0.19}_{-0.20}$& -& $10.9^{+0.4}_{-0.5}$& $2.83^{+1.17}_{-1.41}$& $0.80^{+0.20}_{-0.38}$\\
{\tt IMRPhenomHM (2g)} & $58.2^{+10.5}_{-10.1}$& $29.7^{+7.1}_{-7.4}$& $87.6^{+12.7}_{-11.2}$& $0.51^{+0.17}_{-0.16}$& $0.30^{+0.18}_{-0.21}$& -& $11.1^{+0.4}_{-0.4}$& $2.18^{+1.18}_{-1.05}$& $0.69^{+0.31}_{-0.43}$\\
{\tt SEOBNRv4 (1g)} & $44.3^{+8.5}_{-7.2}$& $38.7^{+7.0}_{-6.9}$& $83.0^{+14.2}_{-11.6}$& $0.88^{+0.12}_{-0.14}$& $0.39^{+0.21}_{-0.20}$& -& $10.8^{+0.4}_{-0.4}$& $2.98^{+1.34}_{-1.50}$& $0.79^{+0.21}_{-0.45}$\\
{\tt SEOBNRv4 (2g)} & $53.9^{+11.5}_{-11.1}$& $31.2^{+8.3}_{-7.5}$& $85.1^{+14.3}_{-11.4}$& $0.57^{+0.21}_{-0.20}$& $0.34^{+0.19}_{-0.20}$& -& $10.8^{+0.4}_{-0.4}$& $2.68^{+1.28}_{-1.36}$& $0.78^{+0.22}_{-0.46}$\\
{\tt SEOBNRv4HM (1g)} & $45.1^{+8.4}_{-7.7}$& $38.5^{+7.8}_{-6.3}$& $83.7^{+14.3}_{-12.0}$& $0.87^{+0.13}_{-0.14}$& $0.38^{+0.20}_{-0.19}$& -& $10.9^{+0.4}_{-0.4}$& $2.94^{+1.35}_{-1.43}$& $0.78^{+0.22}_{-0.47}$\\
{\tt SEOBNRv4HM (2g)} & $56.6^{+9.3}_{-9.9}$& $30.4^{+6.2}_{-7.7}$& $86.8^{+11.5}_{-11.0}$& $0.54^{+0.16}_{-0.16}$& $0.29^{+0.19}_{-0.18}$& -& $11.1^{+0.4}_{-0.4}$& $2.43^{+1.21}_{-1.04}$& $0.72^{+0.28}_{-0.42}$\\
\end{tabular}
\caption{Parameters of GW170729 obtained with various waveform models and the 1g and 2g priors. 
We quote median values and 90\% credible intervals for the primary mass, the secondary mass, the total mass, the effective spin $\chi_{\rm{eff}}$, and the effective precession parameter $\chi_{\rm{p}}$. 
For the mass ratio we quote the median value and the 90\% HPD. 
All masses are given in the source frame. 
The effective precession parameter $\chi_{\rm{p}}$ is absent in the spin-aligned models.
}
\label{table:par-table-2g}
\end{table*}

\section{Morphology-Independent analysis}
\label{BWresults}

\begin{figure*}[]
\includegraphics[width=\columnwidth,clip=true]{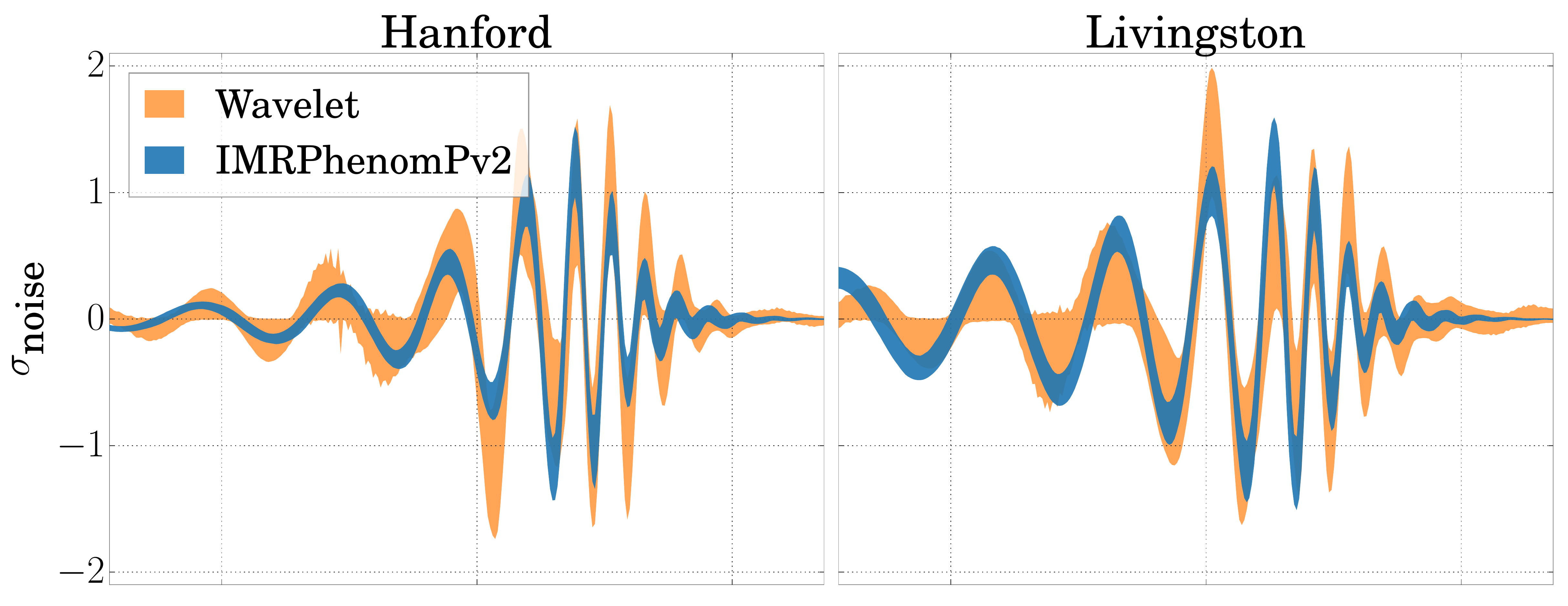}
\includegraphics[width=\columnwidth,clip=true]{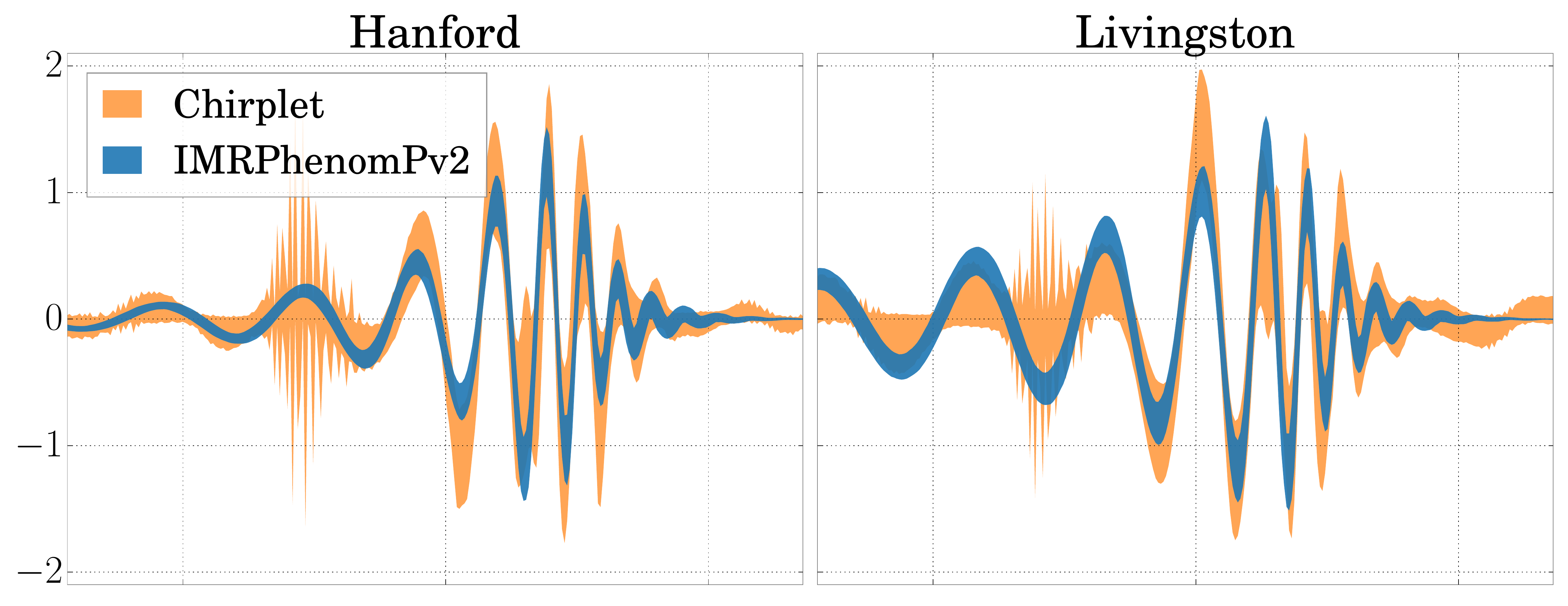}
\includegraphics[width=\columnwidth,clip=true]{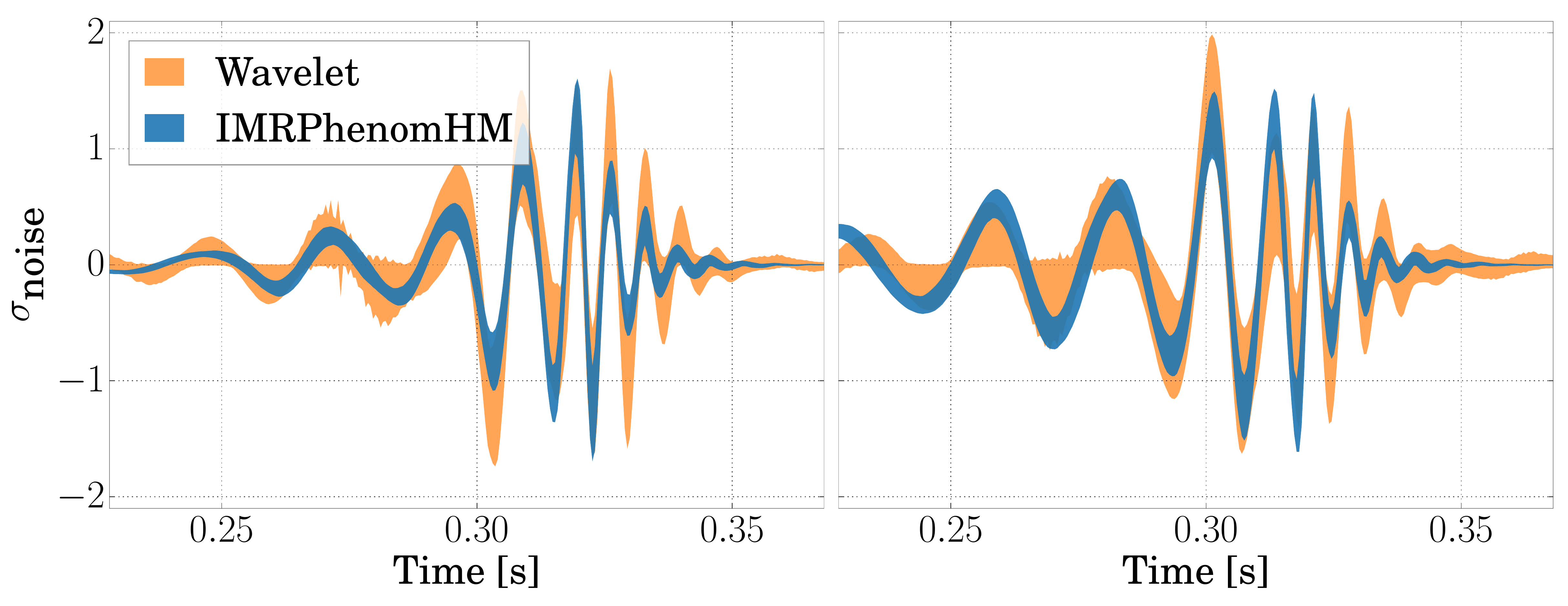}
\includegraphics[width=\columnwidth,clip=true]{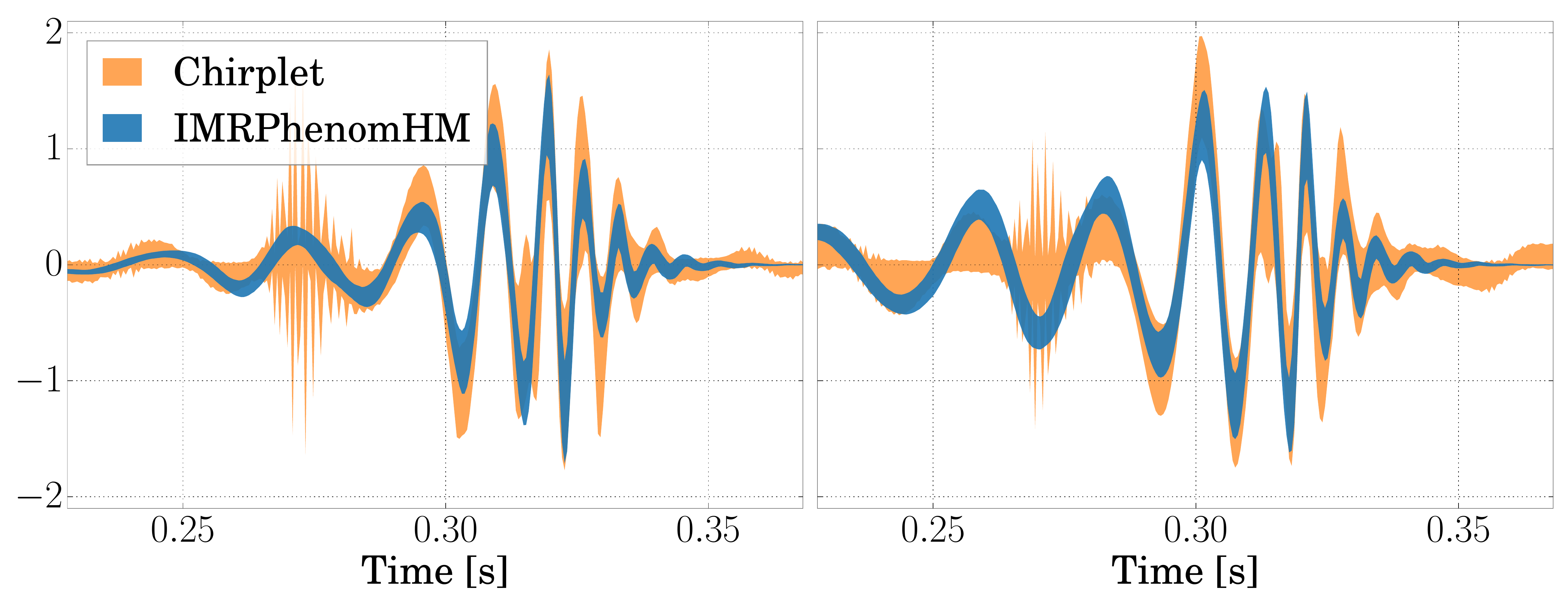}
\caption{Signal reconstruction comparison for GW170729. 
We plot the 90\% credible interval of the whitened strain data as a function of time for each of the LIGO detectors computed with CBC waveform models (blue) and {\tt BayesWave} (orange). 
The top plots show {\tt IMRPhenomPv2} while the bottom plots show {\tt IMRPhenomHM}. The left plots use the wavelet model of {\tt BayesWave}, while the right plots use the chirplet model. The $\textrm{x}$ axis represents the time in seconds from the nearest integer GPS time before the event. The $\textrm{y}$ axis represents the strain amplitude whitened using a filter which is the inverse Amplitude Spectral Density (ASD) of the noise in the detector. The units are in multiples of the standard deviation of the noise.
The generic signal reconstruction is consistent with the CBC signal reconstruction both when the latter includes higher-order modes and when it does not.}
\label{fig:bayeswave}
\end{figure*}

The studies presented in the previous section relied on specific waveform models for the signal emitted during a CBC as predicted by GR. 
We here follow a more generic approach and analyze GW170729 in a morphology-independent way that does not explicitly assume it is a CBC described by the currently available waveform models. 
We use {\tt BayesWave} to reconstruct the signal and then compare this reconstruction to the one obtained with CBC models. 
The comparison is shown in Fig.~\ref{fig:bayeswave}, where we plot the whitened strain as a function of time. 
At each time, the shaded region denotes the 90\% credible interval of the reconstruction using CBC waveform models (blue) and {\tt BayesWave} (orange). 
The left panel is obtained with the wavelets, while the right panel is obtained with chirplets. 
The top row is made with the CBC waveform model {\tt IMRPhenomPv2} which includes the effect of spin precession, but not higher-order modes. 
The bottom row uses {\tt IMRPhenomHM} which assumes that the spins remain aligned with the orbital angular momentum, but includes higher-order modes.

As discussed in~\cite{LIGOScientific:2018mvr}, {\tt BayesWave} can sometimes reconstruct features that are not present in the CBC reconstructions as can be seen in the right hand side panels of Fig.~\ref{fig:bayeswave}, for example around t=0.27s on the right panel. 
Unlike CBC model waveforms, wavelet-based models are not limited to a physically motivated waveform morphology. 
As a result, the {\tt BayesWave} sampler can sometimes pick up random coherence between nearby noise samples. 
However, these outlying wavelets do not point to any potential additional features in the waveform. 
In fact, they are absent in the 50\% credible intervals of the reconstruction, implying that they have a low significance. 
Similar outliers were observed in {\tt BayesWave} analyses applied to simulated signals added to real data.

We find broad agreement between the CBC reconstruction and the {\tt BayesWave} reconstruction in all cases.
In particular, the 90\% credible intervals obtained with the two methods overlap for all waveform models and {\tt BayesWave} basis functions. 
The agreement suggests that the omission of higher-order modes does not degrade the reconstruction enough to leave a coherent residual detectable 
by {\tt BayesWave}. 
This conclusion is in agreement with the results of the previous section, as well as the corresponding reconstruction plot in~\cite{LIGOScientific:2018mvr}.

\begin{figure*}[]
\includegraphics[width=\columnwidth,clip=true]{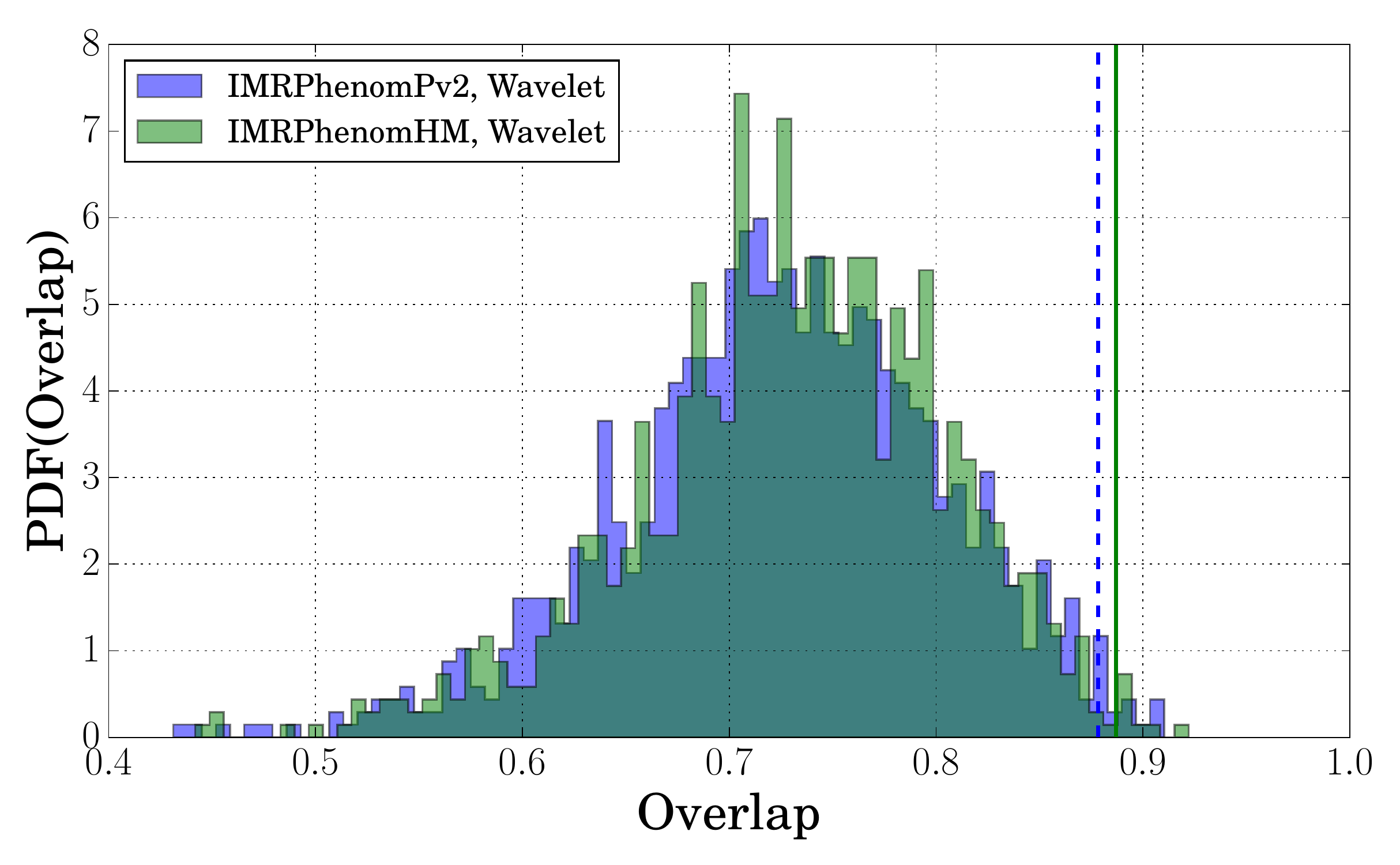}
\includegraphics[width=\columnwidth,clip=true]{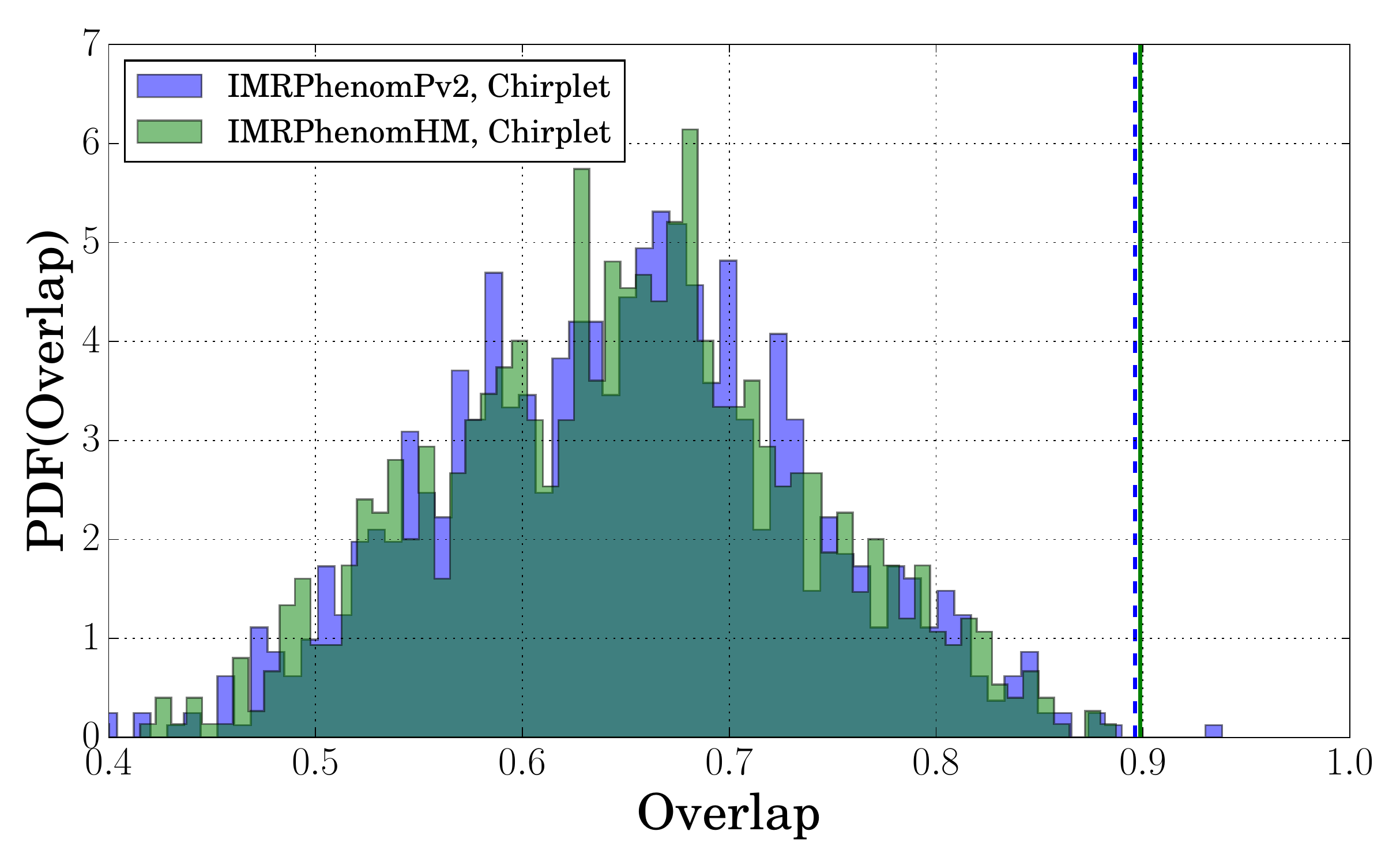}
\caption{Overlap between the {\tt BayesWave} and the CBC reconstruction for GW170729. We plot the overlap histogram between 1000 random {\tt BayesWave} waveform samples and the ML CBC waveform obtained with {\tt IMRPhenomPv2} (blue) and {\tt IMRPhenomHM} (green). 
The left panel uses {\tt BayesWave}'s wavelet model, while the right panel uses the chirplet model. 
The solid and dashed vertical lines represent the overlap of the MBW reconstruction with the ML CBC waveform, with the blue dashed and solid green lines representing the overlaps computed using {\tt IMRPhenomPv2} and {\tt IMRPhenomHM} respectively. 
As is described in the text, we expect these overlaps to be higher than those computed using individual {\tt BayesWave} samples.}
\label{fig:overlap}
\end{figure*}

In order to make this statement quantitative, we draw 1000 samples from the {\tt BayesWave} posterior and compute their overlap with the maximum likelihood (ML) CBC waveform from the analyses using {\tt IMRPhenomPv2} and {\tt IMRPhenomHM}.
The posterior distribution for the overlap is given in Fig.~\ref{fig:overlap} for wavelets (left) and chirplets (right). 
We find overlaps typically between $0.6-0.9$. This large spread in the overlaps is a result of the inherent flexibility in wavelet-based analyses causing a large variance in the reconstructed signal morphology. 
Therefore, unlike CBC waveform samples which are motivated by a physical theory, a single {\tt BayesWave} sample is not constrained by physical reconstruction considerations, other than propagation at the speed of light. 
Instead, the median {\tt BayesWave} waveform (MBW), defined as being the median across the sample waveforms at each time or frequency step, represents a collective estimate across samples, assumed to represent ``the wisdom of the crowd". 
The MBW is a stable estimate of the reconstruction as it is relatively immune to the stochastic fluctuations in the variable dimensional sampler. 
Each of the four vertical lines in Fig.~\ref{fig:overlap} represent the overlap values between the MBW and the ML CBC waveform. 
They are summarized in Table~\ref{table:overlaps} where we also show results obtained using {\tt SEOBNR} waveform models. 
The ML CBC waveform has in general a higher agreement with the MBW waveform than with each of the individual samples.

\begin{table}[]
\centering
  \begin{tabular}{l |c c}
    \hline 
       &Wavelets&Chirplets\\ \hline 
       {\tt IMRPhenomPv2}&$0.88$&$0.90$\\
       {\tt IMRPhenomD}&$0.87$&$0.89$\\
       {\tt IMRPhenomHM}&$0.89$&$0.90$\\
       {\tt SEOBNRv4}&$0.87$&$0.90$\\       
       {\tt SEOBNRv4HM}&$0.88$&$0.89$
   \end{tabular}
   \caption{Overlaps between the median {\tt BayesWave} reconstruction and the maximum likelihood CBC waveform with different waveform models.}
   \label{table:overlaps}
\end{table}

We find that waveforms both with and without higher-order modes achieve large overlaps with the MBW reconstruction, in the range of $0.87-0.9$, consistent with expectations. 
In fact, Ref.~\cite{TheLIGOScientific:2016uux} showed that for masses and SNRs typical of GW170729 the expected overlap between simulated signals and their median reconstructions is in the $0.85-0.9$ range at the 1-$\sigma$ level, similar to what we obtain here. 
The small remaining disagreement between the ML CBC reconstruction and the MBW reconstruction is due to the fact that, unlike modeled analyses, {\tt BayesWave} is only sensitive to excess signal that stands out and above the detector noise. 
This means that it is less sensitive than CBC analyses in the lower frequencies, $<40$ Hz.
We find that if we increase the low frequency cut off in the overlap calculation to $40$ Hz, the overlaps improve by $\sim 0.07$ for each pair of {\tt BayesWave} and CBC waveforms.

Besides the good reconstruction, we find that waveforms with higher-order modes lead to similar overlaps with the {\tt BayesWave} reconstruction to waveforms without higher-order modes. 
We perform the Kolmogorov--Smirnov test for the two overlap distributions for each panel of Fig.~\ref{fig:overlap} and find only 0.048 and 0.017, which implies that both {\tt IMRPhenomPv2} and {\tt IMRPhenomHM} reconstruct the data comparably well. This confirms that GW170729 is consistent with a CBC and that the higher-order modes are not strong enough to lead to a degradation of the signal reconstruction if neglected.

\section{Conclusions}
\label{conclusions}

We analyze the publicly available strain data for GW170729, the highest mass and most distant confirmed GW detection by the LIGO and Virgo detectors. 
In particular we investigate the effect of higher-order modes and spin priors on the inference of the source parameters. 
We find that higher-order modes leave small but noticeable effects, while spin priors affect the spin measurements as anticipated.

We find that the inclusion of higher-order modes in the models leads to changes in the estimates for the mass ratio, the effective spin, and the SNR. 
Our updated parameter measurements imply decreased support for equal binary component masses and nonzero effective spin. 
In particular we conclude that the mass ratio is $(0.3-0.8)$ at the 90\% credible level, a value that excludes equal masses. 
We also find that the 90\% credible interval for the effective spin parameter has changed from $(0.11- 0.58)$ as reported in~\cite{LIGOScientific:2018mvr} to $(-0.01-0.50)$, which now marginally includes zero. 
The effective spin parameter still has a 94\% probability of being positive.

Consistent with these findings, we compute the BF in favor of the presence of higher-order modes, and find it to be 5.1:1. 
Moreover, their omission does not dramatically change the measured parameters, which would happen if they were strong \cite{Varma:2016dnf,Bustillo:2015qty,Graff:2015bba}. 
This conclusion is also consistent with the fact that current matched-filter searches for CBCs have a reduced efficiency toward signals with strong higher harmonics \cite{Harry:2017weg,Capano:2013raa} and this event was indeed observed in both the GstLAL and PyCBC searches.

We argue that the observed changes in parameter measurements are not due to systematic errors in the CBC waveform models. 
We compare results obtained with different waveform models with and without higher-order modes, 
including waveforms computed with NR. 
That leads us to believe that both the increased support for unequal masses and the decreased support for nonzero spin are robust conclusions.

We emphasize that the fact that the evidence for higher-order modes is weak does not contradict the fact that waveforms with higher-order modes lead
to improved parameter measurements. In fact, accurate modeling of relevant physical effects can improve parameter measurements.
This can be because said physical effect is present even if it is too weak to unequivocally detect, or because it helps exclude regions of the parameter space for which that effect
would be larger than what we observe. Similar shifts in the posteriors (though in the opposite direction for the distance and inclination)
where in fact observed in~\cite{Kumar:2018hml} when reanalyzing GW150914 with waveforms that include higher-order modes.

We augment the analysis using CBC waveform models with a morphology-independent analysis using {\tt BayesWave}. 
We find broad agreement between the CBC analysis and the generic analyses regardless of whether the CBC model uses higher-order modes or not. 
We quantify this conclusion in terms of the overlap between the CBC and the generic reconstruction, which we find to be $\sim 0.9$, consistent with expectations for signals of this mass and SNR~\cite{TheLIGOScientific:2016uux}.

Posterior samples from all analyses are available at~\cite{170729SamplesRelease}.

\acknowledgements
 
We thank Christopher Berry, Thomas Dent, Tristano DiGirolamo, Bhooshan Gadre, Roland Haas, Nathan Johnson-McDaniel, Riccardo Sturani, and Aaron Zimmerman for helpful comments and suggestions.
This research has made use of data, software and/or web tools obtained from the Gravitational Wave Open Science Center (https://www.gw-openscience.org), a service of LIGO Laboratory, the LIGO Scientific Collaboration and the Virgo Collaboration. LIGO is funded by the U.S. National Science Foundation. Virgo is funded by the French Centre National de Recherche Scientifique (CNRS), the Italian Istituto Nazionale della Fisica Nucleare (INFN) and the Dutch Nikhef, with contributions by Polish and Hungarian institutes.
The Flatiron Institute is supported by the Simons Foundation.
JCB acknowledges support from Australian Research Council Discovery Project DP180103155.
CJH, KKYN and SV acknowledges the support of the National Science Foundation and the LIGO Laboratory.
LIGO was constructed by the California Institute of Technology and Massachusetts Institute of 
Technology with funding from the National Science Foundation and operates under cooperative agreement PHY-1764464.
MH was supported by Science and Technology Facilities Council (STFC) grant ST/L000962/1 and European Research Council Consolidator Grant 647839.
Parts of this research were conducted by the Australian Research Council Centre of Excellence for Gravitational Wave Discovery (OzGrav), through project number CE170100004.
ROS and JL are supported by National Science Foundation (NSF) PHY-1707965 and PHY-1607520.
PS acknowledges support from the NWO Veni grant no. 680-47-460 and the Science and Technology Facilities Council grant ST/N000633/1.
MH acknowledges support from the Swiss National Science Foundation (SNSF) Grant IZCOZ0\_177057.
LS was supported by the Young Elite Scientists Sponsorship Program by the China Association for Science and Technology (2018QNRC001), and partially supported by the National Science Foundation of China (11721303), and XDB23010200.
The GT authors gratefully acknowledge the NSF for financial support from Grants No. PHY 1806580, PHY 1809572, and TG-PHY120016. Computational resources were provided by XSEDE and the Georgia Tech Cygnus Cluster.
The RIT authors gratefully acknowledge the NSF for financial support from Grants No. PHY-1607520,
No. PHY-1707946, No. ACI-1550436, No. AST-1516150,
No. ACI-1516125, No. PHY-1726215.  This work used the Extreme
Science and Engineering Discovery Environment (XSEDE) [allocation
  TG-PHY060027N], which is supported by NSF grant No. ACI-1548562.
Computational resources were also provided by the NewHorizons, BlueSky
Clusters, and Green Prairies at the Rochester Institute of Technology,
which were supported by NSF grants No.\ PHY-0722703, No.\ DMS-0820923,
No.\ AST-1028087, No.\ PHY-1229173, and No.\ PHY-1726215.
The SXS authors at Caltech acknowledge the Sherman Fairchild Foundation, and NSF grants
PHY-1708212 and PHY-1708213.
GL, NA, and AG are supported by NSF PHY-1606522, PHY-1654359, PHY-1654359. Computations were done using the orca cluster supported in part by PHY-1429873 and by Cal State Fullerton. AG is also supported in part by Nancy Goodhue-McWilliams.
The authors are grateful for computational resources provided by the LIGO Laboratory and supported by National Science Foundation Grants PHY-0757058 and PHY-0823459. 
Some of the computational work for this manuscript was also carried out on the computer cluster {\tt Vulcan} at the Max Planck Institute for Gravitational Physics in Potsdam.  
Plots have been made with {\tt matplotlib}~\cite{Hunter:2007} and {\tt corner}~\cite{corner}.

\appendix

\section{Differences between {\tt LALInference} and {\tt RIFT}}
\label{RIFTtroubles}

\begin{table}[]
\centering
  \begin{tabular}{cccccc}
    \hline \hline
Parameter    &$q$ &$\chi_{\rm{eff}}$ \\ 
\hline
{\tt SEOBNRv4HM}  ($\chi$) &$0.59^{+0.34}_{-0.24}$& $0.29^{+0.25}_{-0.30}$ \\
NR/\NRSur{}   ($\chi$) & $0.64^{+0.32}_{-0.24}$& $0.33^{+0.22}_{-0.26}$ \\
NR/\NRSur{}   ($\chi_z$) & $0.67^{+0.30}_{-0.24}$& $0.40^{+0.22}_{-0.24}$ \\
NR/\NRSur{} HM  ($\chi$) & $0.58^{+0.34}_{-0.24}$& $0.29^{+0.23}_{-0.28}$ \\
NR/\NRSur{} HM  ($\chi_z$) & $0.62^{+0.32}_{-0.23}$& $0.37^{+0.25}_{-0.24}$\\
NR/\NRSur{}   (1g) &  $0.91^{+0.08}_{-0.12}$& $0.38^{+0.20}_{-0.25}$ \\
NR/\NRSur{}   (2g)&$0.58^{+0.23}_{-0.20}$& $0.34^{+0.19}_{-0.22}$ \\
NR/\NRSur{} HM  (1g) & $0.91^{+0.09}_{-0.12}$& $0.38^{+0.21}_{-0.23}$\\
NR/\NRSur{} HM  (2g) &$0.55^{+0.21}_{-0.19}$& $0.31^{+0.21}_{-0.23}$ \\
\end{tabular}
\caption{Estimates for the parameters of GW170729 obtained with {\tt RIFT} using various priors and waveform models. 
We quote median values and 90\% credible intervals for the the effective spin and HPD for the mass ratio. 
We follow similar notation as Tables~\ref{table:par-table} and~\ref{table:par-table-2g}.
}
\label{table:par-table_RIFT}
\end{table}

In Sec.~\ref{2HM} we discuss how higher-order modes affect the posterior distributions for the various source properties of GW170729.
We argue that waveform systematics are small since results with {\tt IMRPhenomHM} and {\tt SEOBNRv4HM} agree well with each other,
and the latter agrees well with NR waveform models. The investigation of waveform systematics, though, also reveals that there is a residual
small disagreement between results with {\tt SEOBNRv4HM} obtained with {\tt LALInference} and {\tt RIFT}. We have performed extensive 
investigations into the nature of this disagreement and have been ultimately unable to pinpoint its origin.

{\tt LALInference} and {\tt RIFT} are independently-implemented codes with differences in data-handling, likelihood estimation, algorithm, etc.
Despite these differences, in this work we have found good agreement between results obtained by the two algorithms for waveform approximants
that do not include higher-order modes, see Tables~\ref{table:par-table} and~\ref{table:par-table_RIFT}. However, for waveforms with higher-order modes and in particular the 
direct comparison using {\tt SEOBNRv4HM}, we find that the two codes produce results that differ for the mass ratio at the $7\%$ level. We also 
find that the two codes produce consistent results for the effective spin and the detector-frame total mass of GW170729, though
we are unable to check the source-frame total mass which {\tt RIFT} did not compute. See Table~\ref{table:par-table_RIFT} for more estimates.

We performed a number of reanalyses of the data in order to test the effects of various differences between the two algorithms.
On the {\tt RIFT} side these tests include: the NR grid, 
the specific choice of fitting coordinates, the noise PSD calculation, the data handling, the sampling rate, 
the lower frequency cut-off, the Monte-Carlo integration, the likelihood evaluation, the summation of higher-order modes to get the waveform, and the time window of the analysis. 
More technical details about these tests are presented in~\cite{Lange:2018}.
We also performed {\tt LALInference} runs ignoring the detector calibration uncertainty.
We found that none of these tests could account for the shift in the mass ratio posteriors.

Given that and the long history of testing and usage of {\tt LALInference}, in this paper we also follow previous studies by the LIGO/Virgo 
Collaborations, for example~\cite{Abbott:2018wiz,LIGOScientific:2018mvr}, and use {\tt LALInference}
 for our main results. We do note, though, that {\tt RIFT} results are qualitatively consistent and quantitatively close 
 to {\tt LALInference} and the discrepancy is only noticeable when higher-order modes are taken into account. 
 The small residual disagreement will be the focus of future investigations.

\bibliography{OurRefs}

\end{document}